\documentclass[a4paper,11pt]{article}
\pdfoutput=1 
\usepackage{mathrsfs, amssymb, amsmath}  
\usepackage{jcappub} 

\usepackage[T1]{fontenc} 

\usepackage{epsfig, cancel}
\usepackage{latexsym}
\usepackage{natbib, comment}
\usepackage{url}
\usepackage{dcolumn}
\usepackage{multirow}
\usepackage{color}
\usepackage{cancel}
\usepackage{soul}
\usepackage[normalem]{ulem}
\usepackage{amsfonts,amssymb,amsmath, txfonts}
\usepackage{graphicx,epsfig}
\usepackage{psfrag}
\usepackage{hyperref}
\hypersetup{colorlinks=true}
\usepackage{mathtools}
\usepackage{enumitem}
\usepackage{float}
\usepackage[dvipsnames]{xcolor}
\usepackage{xcolor}
\hypersetup{ linktoc=all,
    colorlinks, linkcolor={brightpink},
    citecolor={blue}, urlcolor={blue}
}
\definecolor{rosy}{RGB}{230,235,252}
\definecolor{myframetitle}{RGB}{90,89,170}
\definecolor{myblocktitle}{RGB}{140,185,249}
\definecolor{mytitle}{RGB}{10,80,26}

\definecolor{darkgreen}{RGB}{27,130,45}
\definecolor{darkblue}{rgb}{0,0,0.3}
\definecolor{darkred}{rgb}{0.7,0,0}

\definecolor{light gray}{RGB}{220,220,220}
\definecolor{dark purple}{RGB}{108,0,217}
\definecolor{pink}{RGB}{190,20,100}
\definecolor{orang}{RGB}{193,63,0}
\definecolor{green}{RGB}{11,98,17}
\definecolor{darkpink}{RGB}{153,0,76}
\definecolor{bluegreen}{RGB}{0,102,102}
\definecolor{greenlagan}{RGB}{0,102,0}
\definecolor{redgreen}{RGB}{102,102,0}
\definecolor{Redgreen}{RGB}{153,76,0}
\definecolor{vividviolet}{rgb}{0.62, 0.0, 1.0}
\definecolor{amaranth}{rgb}{0.9, 0.17, 0.31}
\definecolor{palatinateblue}{rgb}{0.15, 0.23, 0.89}
\definecolor{brightpink}{rgb}{1.0, 0.0, 0.5}
\definecolor{cornflowerblue}{rgb}{0.39, 0.58, 0.93}
\definecolor{deepcarminepink}{rgb}{0.94, 0.19, 0.22}
\definecolor{radicalred}{rgb}{1.0, 0.21, 0.37}

\def\H0{{\text{H}\hspace*{-2.05mm}\text{H} 0\hspace*{-1.35mm}0\ }}

\def\be{\begin{equation}}
\def\ee{\end{equation}}
\def\beq{\begin{equation}}
\def\eeq{\end{equation}}
\def\bea{\begin{eqnarray}}
\def\eea{\end{eqnarray}}
\newcommand{\dd}{\textrm{d}}
\newcommand{\nn}{\nonumber \\}

\title{\centerline{Is Local $H_0$ At Odds With Dark Energy EFT?}}

\author[a,b]{Bum-Hoon Lee}
\author[a]{, Wonwoo Lee}
\author[a,b]{, Eoin \'O Colg\'ain}
\author[c]{, M. M. Sheikh-Jabbari}
\author[a,b]{and Somyadip Thakur}

\affiliation{\vskip 3mm $^a$ Center for Quantum Spacetime, Sogang University, Seoul 121-742, Korea}
\affiliation{$^b$ Department of Physics, Sogang University, Seoul 121-742, Korea}
\affiliation{$^c$ School of Physics, Institute for Research in Fundamental Sciences (IPM), P.O.Box 19395-5531, Tehran, Iran
\vskip 3mm }

\emailAdd{bhl@sogang.ac.kr, warrior@sogang.ac.kr, eoin@sogang.ac.kr, jabbari@theory.ipm.ac.ir, somyadip@sogang.ac.kr}

\abstract{
Local $H_0$ determinations currently fall in a window between $H_0 \sim 70$ km/s/Mpc (TRGB) and $H_0 \sim 76$ km/s/Mpc (Tully-Fisher). In contrast, BAO data calibrated in an early $\Lambda$CDM universe are largely consistent with Planck-$\Lambda$CDM, $H_0 \sim 67.5$ km/s/Mpc. Employing a generic two parameter family of evolving  equations of state (EoS) for dark energy (DE) $w_{\textrm{DE}}(z)$ and mock BAO data, we demonstrate that if i) $w_{\textrm{DE}}(z=0) < -1$ and ii) integrated DE density less than $\Lambda$CDM, then $H_0$ increases. EoS that violate these conditions at best lead to modest $H_0$ increases within $1 \sigma$. Tellingly, Quintessence and K-essence satisfy neither condition, whereas coupled Quintessence can only satisfy ii). Beyond these seminal DE Effective Field Theories (EFTs), we turn to explicit examples. Working model agnostically in an expansion in powers of redshift $z$, we show that Brans-Dicke/$f(R)$ and Kinetic Gravity Braiding models within the Horndeski  class can lead to marginal and modest increases in $H_0$, respectively. We confirm that as far as increasing $H_0$ is concerned, no DE EFT model can outperform the phenomenological two parameter 
family of the DE models. Evidently, the late universe may no longer be large enough to accommodate $H_0$, BAO and DE described by EFT.  
}

\begin{document}
\maketitle
\flushbottom

\section{Introduction} 
Field theory has emerged as the leading framework to formulate physics, from early universe cosmology and particle physics to condensed matter and biophysics. Not only is this true classically, whether one considers Maxwell's electromagnetism or Einstein's General Theory of Relativity (GR), but Quantum Electrodynamics (QED) has been stringently tested.\footnote{The fine structure constant $\alpha$ is known to a precision of better than a part in a billion or a relative uncertainty of $7.2 \times 10^{-10}$ \cite{Hanneke:2008tm}. In contrast, the relative uncertainty in Newton's constant $G$ is $2.2 \times 10^{-5}$ according to CODATA 2018.}  In contrast, the concordance $\Lambda$CDM model is a phenomenological model in which about 95\% of the energy budget of the universe is in the dark sector. While there are various field theory models to describe the dark matter sector, the remaining 69\% in the dark energy (DE) sector is described by the cosmological constant $\Lambda$, which is simply a \textit{placeholder for missing physics}. Common lore states that $\Lambda$ not only suffers from  cosmological constant problems \cite{Weinberg:1988cp}, but also a ``coincidence problem'' \cite{coincidence, Zlatev:1998tr}. These problems have historically served as the motivation to replace $\Lambda$ with an Effective Field Theory (EFT) description, typically captured by additional dynamical scalar fields. This has motivated the class of scalar-tensor field theories, which govern the gravity and DE sector of the cosmological model, most notably Quintessence \cite{Ratra:1987rm,Wetterich:1987fm} and K-essence models \cite{Armendariz-Picon:1999hyi, Chiba:1999ka, Armendariz-Picon:2000ulo}. However, the cosmological constant problem remains.   

\begin{figure}[htb]
\centering
\hspace*{-10mm} \includegraphics[width=90mm]{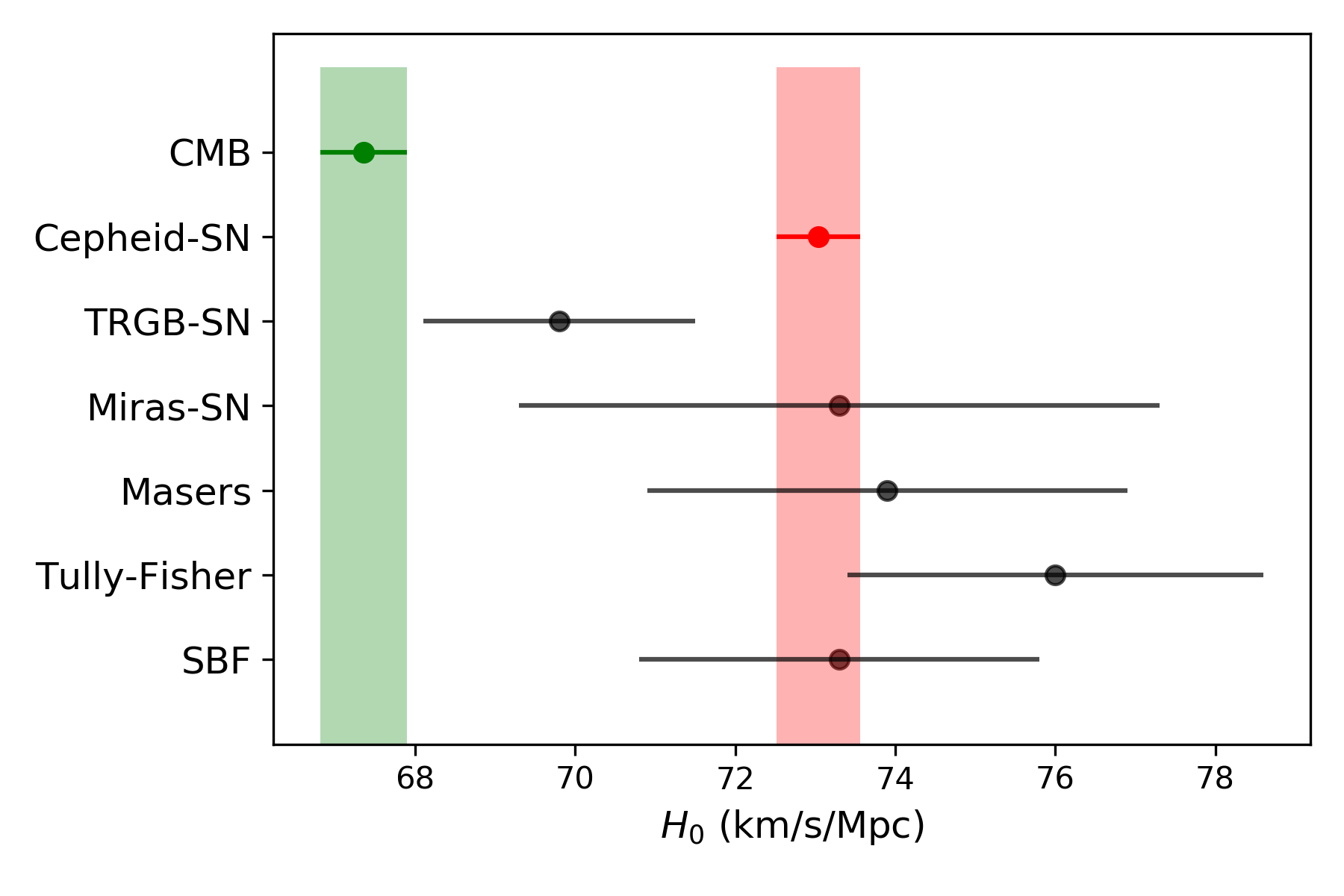}
\caption{The Planck $H_0$ determination alongside local determinations based on Cepheids \cite{Riess:2021jrx}, TRGB \cite{Freedman:2021ahq}, Miras \cite{Pesce:2020xfe}, Masers \cite{Pesce:2020xfe}, Tully-Fisher \cite{Kourkchi:2020iyz} and SBF \cite{Blakeslee:2021rqi}. TRGB and Tully-Fisher determinations more or less bound local determinations. The coloured regions document a $\sim 5 \sigma$ disagreement.}
\label{fig:H0}
\end{figure}

In recent years local determinations of $H_0$ across different distance indicators, including Cepheids \cite{Riess:2021jrx}, Tip of the Red Giant Branch (TRB) \cite{Freedman:2021ahq}, water megamasers \cite{Pesce:2020xfe}, Tully-Fisher relation \cite{Kourkchi:2020iyz} and surface brightness fluctuations (SBF) \cite{Blakeslee:2021rqi} have all returned values \textit{larger} than Planck-$\Lambda$CDM \cite{Planck:2018vyg}. While it is true that some determinations of $H_0$ in the late universe recover values consistent with Planck-$\Lambda$CDM, notably strong lensing time delay \cite{Birrer:2020tax} and gravitational waves \cite{LIGOScientific:2017adf} (see more recently \cite{Palmese:2021mjm}), strictly speaking, these are not local determinations, since they must assume a cosmological model. Furthermore, they typically have larger errors than direct local measurements and that of Planck-$\Lambda$CDM. There are ongoing debates about whether a tension exists, and even if it does, whether it is of significance or not, but it is difficult to imagine local $H_0$ values converging to a value below Planck-$\Lambda$CDM.\footnote{Observe also that the base Planck-$\Lambda$CDM model has fixed (unrealistic) values for the neutrino masses. Relaxing this constraint, the central value of $H_0$ is even lower.} The current state of the art is encapsulated in Fig. \ref{fig:H0}, where we have only highlighted some indicative local determinations across observables. For a comprehensive overview of recent results, we refer to the reader to Fig. 1 of \cite{DiValentino:2021izs}, where it is clear that TRGB at $H_0 \sim 70$ km/s/Mpc and Tully-Fisher at $H_0 \sim 76$ km/s/Mpc constitute \textit{outliers}.\footnote{Some recent obsession with outliers in the literature reminds us that science is an endeavour pursued by humans.}

This biasing of local $H_0$ determinations to larger values can be a game changer for the traditional DE paradigm. Recall that in the traditional setup, dark matter and DE do not talk to each other. Nevertheless, in this minimal setting, one can replace $\Lambda$ and its equation of state (EoS), $w_{\textrm{DE}}=-1$, with a constant parameter, $w_{\textrm{DE}} = w$, or the two parameter family 
$ w_{\textrm{DE}}  = w_0 + w_1 f(z)$ where $f(z)$ is an arbitrary function of the redshift $z$ such that $f(0)=0$ and $f(z\gg 1)$ remains finite. See \cite{Colgain:2021pmf} for more discussion and analysis. {Within this setting, the prevailing Chevallier-Polarski-Linder (CPL) model \cite{Chevallier:2000qy, Linder:2002et} corresponds to $f(z)=z/(1+z)$. For the CPL  parametrisation it has been shown that restricting to the $w_{\textrm{DE}} > -1$ range of parameter space lowers $H_0$ \cite{Vagnozzi:2018jhn, Vagnozzi:2019ezj, Alestas:2020mvb}}. \textit{This precipitates the notion that  DE models  satisfying the Null Energy Condition, $1+w_{\textrm{DE}} \geq 0$, which encompass a large class of Effective Field Theories (EFTs) (but, by no means all!), cannot perform better than $\Lambda$ when it comes to recovering local $H_0$ determinations}. In other words, if one is left chasing a higher local $H_0$ within EFT, it may be best to do it with a scalar fixed in the minimum of a potential in the late universe and having a dynamical DE EFT with redshift dependent EoS does not help to considerably increase $H_0$. Of course, one may even worry that this makes DE EFT redundant, and if so, then how should one interpret the placeholder $\Lambda$? 

However, it would be hasty to base conclusions solely on the  CPL parametrisation, as it only covers a specific class of Quintessence potentials \cite{Scherrer:2015tra} and is quantifiably less well tailored to the low redshifts where DE is most relevant \cite{Colgain:2021pmf}. One can tackle the more general problem for Quintessence by expanding the dynamical scalar in redshift $z$ below $z= 1$. Since $z$ is a small parameter, perturbation is a valid tool. Then, provided any displacement $\Delta \phi = \phi - \phi_0$ in the scalar from its value today $\phi_0$ is small, i. e. $\Delta \phi < 1$ in some units, one has further computational control to expand the Quintessence potential, $V(\phi) = V_0 + V_1 (\Delta \phi) + V_2 (\Delta \phi)^2 + \dots$. Thus, at low redshift, one can solve the background equations for the Quintessence model in terms of a finite number of constant parameters, scan over those parameters and make statements that hold for models ``close to'' flat $\Lambda$CDM. Adopting this approach, Ref. \cite{Banerjee:2020xcn} constructed a large class of Quintessence models, showing that for representative Type Ia supernovae \cite{Pan-STARRS1:2017jku}, cosmic chronometer \cite{Moresco:2016mzx} and BAO data \cite{Beutler:2011hx, Ross:2014qpa, BOSS:2016wmc}, only models that lower $H_0$ are preferred by data.\footnote{There are concrete Quintessence models that increase $H_0$ while worsening fits through a demonstratable larger $\chi^2$ compared to $\Lambda$CDM \cite{Santos:2021lqh}.} This generalised an earlier statement \cite{Colgain:2019joh} beyond specific Quintessence potentials. Thus, intuition gained from $w$CDM and $w_0 w_a$CDM models, in particular CPL, transfers over wholesale to Quintessence, at least perturbatively.  Throughout the advantage of this perturbative approach is simply that one can work in a model agnostic fashion, albeit in a class of models close to flat $\Lambda$CDM.\footnote{The same can be done through direct data reconstructions \cite{Pogosian:2021mcs, Raveri:2021dbu}, but one requires a theory prior if one probes BAO data with correlations below a certain redshift length scale, otherwise wiggly reconstructions ensue, e. g. \cite{Wang:2018fng}. The reason being that one can always enhance the fit to $H(z_i)$ and $D_{A}(z_i)$ BAO constraints at separated $z_i$ by deviating from strictly increasing functions, such as flat $\Lambda$CDM. In contrast, in Taylor expansion, the analyticity properties of field theories are built in from the outset. See further discussion in \cite{Colgain:2021pmf}.} While resorting to perturbation may sound like a disadvantage, we may have already reached the precision where deviations from flat $\Lambda$CDM can no longer be large and perturbative expansion around it allows us to probe a very large class of models within DE EFT.

Our primary goal here is to extend the {analysis of  Ref. \cite{Banerjee:2020xcn} to more general DE EFTs, such as the Horndeski class \cite{Horndeski:1974wa, Deffayet:2011gz, Kobayashi:2011nu} of scalar-tensor theories}. The motivation comes from the fact that scalar-tensor theories, and more general modified gravity theories, are routinely touted to have cosmological applications (see  \cite{Clifton:2011jh} for a review). Moreover, within Horndeski theories, one can \textit{prima facie} construct models where $w_{\textrm{DE}} < -1$ \cite{DeFelice:2011bh, Matsumoto:2017qil}, a possibility which does not exist within the Quintessence family. The class is defined by four arbitrary functions of a scalar $\phi$ and its kinetic term, and Taylor expansion once again allows one to treat any unknown functions in an agnostic fashion. Our mission is greatly helped by the fact that post GW170817 \cite{LIGOScientific:2017vwq} and the associated gamma-ray burst GRB 170817A \cite{Goldstein:2017mmi, LIGOScientific:2017zic}, one can justify restricting attention to theories where gravitational waves propagate at the speed of light \cite{Lombriser:2015sxa, Creminelli:2017sry, Ezquiaga:2017ekz, Sakstein:2017xjx, Baker:2017hug}.\footnote{Obviously, this can be relaxed, since the source for GW170817 is at $\sim 40$ Mpc, or redshift $z \sim 0.009$. Thus, any constraints only hold in the local universe, but as further events are observed, the window for viable Horndeski models should narrow. Nevertheless, one can find models evading the current bounds \cite{Bayarsaikhan:2020jww}.} This allows one to focus on the simpler class of Lagrangians,  
\be
\label{horndeski}
\mathcal{L} = G_{2}(\phi, X) + G_{3} (\phi, X) \Box \phi + G_{4} (\phi) R, \quad X := - \frac{1}{2} \partial_{\mu} \phi \partial^{\mu} \phi, 
\ee
which already include some interesting subcases, as illustrated in Table \ref{table1}. Here we adopt a $(-,+,+,+)$ signature for the FLRW metric so that $X=\frac12\dot\phi^2 \geq 0$, where \textit{dot} denotes derivative w.r.t comoving time $t$. See \cite{Kase:2018aps} for a review of these solutions in the aftermath of GW170817. 

\begin{table}[htb]
\centering 
\begin{tabular}{c|c}
\rule{0pt}{3ex} Class & $G_{i}(\phi, X)$ \\
\hline
\rule{0pt}{3ex} Quintessence \cite{Ratra:1987rm,Wetterich:1987fm} & $G_{2} = X-V(\phi), G_{3} = 0, G_{4} = \frac{1}{2} M_{\textrm{pl}}^2$ \\
\rule{0pt}{3ex} K-essence \cite{Armendariz-Picon:1999hyi, Chiba:1999ka, Armendariz-Picon:2000ulo} & $G_{2} = G_{2} (\phi, X), G_{3} = 0, G_{4} = \frac{1}{2} M_{\textrm{pl}}^2$ \\
\rule{0pt}{3ex} Brans-Dicke/$f(R)$ gravity \cite{Brans:1961sx, Bergmann:1968ve} & $G_{2} = G_{2} (\phi, X), G_{3} = 0, G_{4} = G_4 (\phi)$ \\
\rule{0pt}{3ex} Kinetic Gravity Braiding \cite{Deffayet:2010qz} & $G_{2} = G_{2} (\phi, X), G_{3} = G_3(\phi, X), G_{4} = \frac{1}{2} M_{\textrm{pl}}^2$ \\
\end{tabular} 
\caption{Various subcases of the Lagrangian (\ref{horndeski}) in the literature.} 
\label{table1} 
\end{table}

As we show in section \ref{sec:warmup}, using mock BAO data based on Planck-$\Lambda$CDM, one can quickly confirm that replacing the cosmological constant $\Lambda$ with either Quintessence or K-essence is expected to result in a lower cosmological inference of $H_0$. Concretely, we show for $(w_0, w_a)$ parametrisations of the EoS that, if $w_{\textrm{DE}}(z=0) = w_0 < -1$ {and} the integrated DE density is less than the analogous $\Lambda$CDM quantity through to the deceleration-acceleration transition redshift, $z_*$, then $H_0$ increases, and \textit{vice versa}. An immediate corollary of this analysis is that any DE model that significantly increases $H_0$ must satisfy both of these conditions. On the flip side, any model with $w_{\textrm{DE}}(z=0) > -1$, but allows $w_{\textrm{DE}}(z) < -1$ at larger $z$, for example coupled Quintessence \cite{Amendola:1999er, Das:2005yj}, is going to have less success in closing the $H_0$ gap evident in Fig. \ref{fig:H0}. Tellingly, current constraints limit $H_0$ inferences within coupled Quintessence below $H_0 = 70$ km/s/Mpc \cite{Gomez-Valent:2020mqn}, even with a local $H_0$ prior. {Two recent papers \cite{Heisenberg:2022lob, Heisenberg:2022gqk} analytically prove that alleviating $H_0$ tension requires $w_{\textrm{DE}}(z) < -1$.} Our analysis in section \ref{sec:warmup} shows that $w_{\textrm{DE}}(z=0)$ is the strongest indicator of $H_0$ behaviour, so one can expect models that lower $w_{\textrm{DE}}$ at $z = 0$ to perform best.  

Having highlighted the obvious $H_0$ problem for Quintessence, K-essence and essentially any DE EFT with $w_{\textrm{DE}}> -1$ at all redshifts (see also \cite{Heisenberg:2022lob, Heisenberg:2022gqk}), in the latter part of this work we turn our attention to more general models in the Horndeski class, which permit $w_{\textrm{DE}}(z) < -1$ \cite{DeFelice:2011bh, Matsumoto:2017qil}. This allows us to test the intuition gained in section \ref{sec:warmup} that $w_{\textrm{DE}}(z=0)$ is most relevant. Throughout, one is always conscious that $(w_0, w_a)$ parametrisations are not truly general \cite{Scherrer:2015tra, Wen:2021bsc}, but that being said, if one works suitably close to $z=0$, one expects little difference. To that end, introducing a non-minimal coupling places one in the Brans-Dicke/$f(R)$ gravity class \cite{Brans:1961sx, Bergmann:1968ve} (see Table \ref{table1}). However, evidence for evolution in the Newton's constant, and therefore $G_{4}(\phi)$, is weak, and on the contrary, one has strong solar system \cite{Hofmann:2018myc} and BBN constraints \cite{Alvey:2019ctk} (see also \cite{Wang:2020bjk,Ballardini:2021evv, Ballardini:2021eox} {and \cite{Marra:2021fvf, Alestas:2021luu, Alestas:2022xxm} and references therein for an alternative view}). On the minimal assumption that the coupling $G_4(\phi)$ evolves linearly with cosmic time, while employing model agnostic techniques for the Quintessence sector, $G_{2}(\phi, X) = X - V(\phi)$, we show in section \ref{sec:nonmin} that any increase in $H_0$ is expected to be negligible. In other words, despite ongoing debate about whether non-minimal couplings can alleviate $H_0$ tension \cite{Rossi:2019lgt, Ballesteros:2020sik, Braglia:2020iik}, in line with  \cite{Wang:2020bjk,Sakr:2021nja}, we find that a non-minimal coupling can only marginally alleviate $H_0$ tension, at least within reasonable assumptions. Concretely, we observe that the class of models largely fails to penetrate into the phantom regime $w_{\textrm{DE}}< -1$ at $z=0$, so it is consistent with our $(w_0, w_a)$ expectations. In contrast, in section  \ref{sec:KGB}, we find that Kinetic Gravity Braiding (KGB) models \cite{Deffayet:2010qz} can alleviate $H_0$ tension in a more meaningful way and trace the higher inferred values of $H_0$ to $w_{\textrm{DE}}(z=0)$. Nevertheless, it should be stressed that KGB is a minimally coupled model, so final $H_0$ inferences are expected to be in line with \cite{Krishnan:2021dyb}, where it was demonstrated that purely late universe resolutions to $H_0$ tension will struggle to exceed $H_0 = 70$ km/s/Mpc. {A like-for-like comparison between KGB and $(w_0, w_a)$ parametrisations demonstrates that $H_0$ displacements are considerably smaller for KGB, so if the latter cannot resolve $H_0$ tension \cite{Yang:2021flj}, then neither can KGB.}

Finally, we remark that at each stage of the analysis,  we make a direct comparison with the flat $\Lambda$CDM model. In other words, our statements are relative statements. One can contemplate altering the BAO scale through Early Dark Energy (EDE) \cite{Poulin:2018cxd, Niedermann:2019olb} or equivalent early universe physics, but this does not change relative statements. This of course presents an intriguing avenue to resurrect Quintessence etc. as viable DE models. That being said, the age of both the universe and astrophysical objects within it bound $H_0 \lesssim 73$ at $2 \sigma$ confidence interval within FLRW cosmology \cite{Krishnan:2021dyb, Vagnozzi:2021tjv}, so even if EDE works which at present this is far from clear \cite{Hill:2020osr,Ivanov:2020ril,DAmico:2020ods, Niedermann:2020qbw, Murgia:2020ryi, Smith:2020rxx, Jedamzik:2020zmd, Lin:2021sfs, Vagnozzi:2021gjh, Herold:2021ksg}, it is plausible that local $H_0$ may still be biased high.\footnote{One can allow for late-time evolving DE with EDE, but late universe observations are largely consistent with $\Lambda$ \cite{Wang:2022jpo}. Interestingly, EDE may also be consistent with a scale invariant spectral index, $n_s \approx 1$ \cite{Takahashi:2021bti}.} 

\noindent
\textbf{Note added:} While we were in the final stages of preparation for the arXiv submission, two papers \cite{Heisenberg:2022lob, Heisenberg:2022gqk} appeared that have good overlap, and of course are compatible with our main results and statements. 

\section{Comparison of $H_0$ from $w_{\textrm{DE}}(z)$ models and $\Lambda$CDM based on mock data}
\label{sec:warmup}
Here we make the case that $H_0$ must always decrease in K-essence models because $w_{\textrm{DE}}(z) > -1$. Our observation includes Quintessence as a special case, where this fact has already been established \cite{Banerjee:2020xcn}. Let us start with a simple hand waving argument before quantifying later through mock data. Recall that in Einstein frame, the general form for the Hubble parameter of a late universe cosmology comprising a matter and DE sector is 
\be
\label{hubble}
H(z)^2  = H_0^2 \left[ (1- \Omega_{m} ) X(z) + \Omega_{m} (1+z)^3 \right],   
\ee
where $\Omega_{m}$ is matter density and $X(z)$, the normalised DE density $X(z) := \rho_{\textrm{DE}}(z)/\rho_{\textrm{DE}}(0)$, may be expressed in terms of the EoS,
\be
X(z) = \exp  \left( 3 \int_0^z \frac{1+w_{\textrm{DE}}(z^{\prime})}{1+z^{\prime}} \dd z^{\prime} \right).  
\ee
While in the flat $\Lambda$CDM model $X(z) = 1$ for all $z$, it should be clear that $X(z)$ varies with $w_{\textrm{DE}}(z)$, and in particular, $w_{\textrm{DE}}(z) > -1$ ($w_{\textrm{DE}}(z) < -1$) for all $z$ implies $X(z) > 1$ ($X(z) < 1$). Between these two extremes, one may consider more general functions $w_{\textrm{DE}}(z)$ that cross the phantom divide, $w_{\textrm{DE}}=-1$.\footnote{See  \cite{DiValentino:2020naf} for a recent study dedicated to this direction.} 
 
Obviously, data only cares about $z$ dependence. Thus, given the Hubble parameter (\ref{hubble}), high redshift data - CMB or equivalent - effectively constrains the combination $H_0^2 \Omega_{m}$, since DE is traditionally assumed to be effective at low redshifts and $X(z)$ cannot grow appreciably with redshift. For this reason, this combination is for all extensive purposes a constant $\kappa \approx H_0^2 \Omega_{m}$.  In practice, it is not a constant, but there is a finite window in which it varies. This allows us to rewrite (\ref{hubble}) as
\be
\label{hubble2}
H(z)^2 \approx H_0^2 X(z) - \kappa X(z) + \kappa (1+z)^3.
\ee
$H(z)$ is fixed by the data, while $X(z)$ can be raised or lowered  through model selection. It should be clear from (\ref{hubble2}) that once the high redshift data fixes $\kappa$ to the allowed window, the only freedom left is in $H_0$. For this reason, one expects $X(z)$ and $H_0$ to be correlated. However, since $X(z)$ is not a constant parameter, but rather a function, one expects this correlation to hold over a redshift range. Thus, it is reasonable to integrate the (normalised) DE density $X(z)$ over a redshift range and check for correlations with $H_0$. That being said, the redshift range is ambiguous, but there is a deceleration-acceleration transition redshift $z_{*}$ that is \textit{universal} to all DE models. This corresponds to the redshift where acceleration vanishes, i.e. 
\be
\label{func}
- (1+z) H'(z) + H(z) = 0 \quad \Rightarrow \quad \Omega_{m} (1+z_{*})^3 + (1-\Omega_{m}) X(z_{*}) \left[ 1 + 3 w_{\textrm{DE}}(z_{*}) \right] = 0. 
\ee
For data consistent with Planck-$\Lambda$CDM, $\Omega_{m} \approx 0.3$, one can expect $z_{*} \approx 0.6$. Nevertheless, for any DE model in the class (\ref{hubble}) a distinct value exists and one can extract it by solving (\ref{func}). This allows us to define $Y(z_*) = \int_0^{z_*} X(z^{\prime}) \dd z^{\prime}$ and any difference relative to flat $\Lambda$CDM in the same range: 
\be
\Delta Y(z_{*}) :=  \int_0^{z_*} [ X(z^{\prime}) -  X_{\Lambda\textrm{CDM}} (z^{\prime}) ] \dd z^{\prime} =  Y(z_*) - z_*.
\ee

\begin{figure}[htb]
  \centering
  \includegraphics[width=90mm]{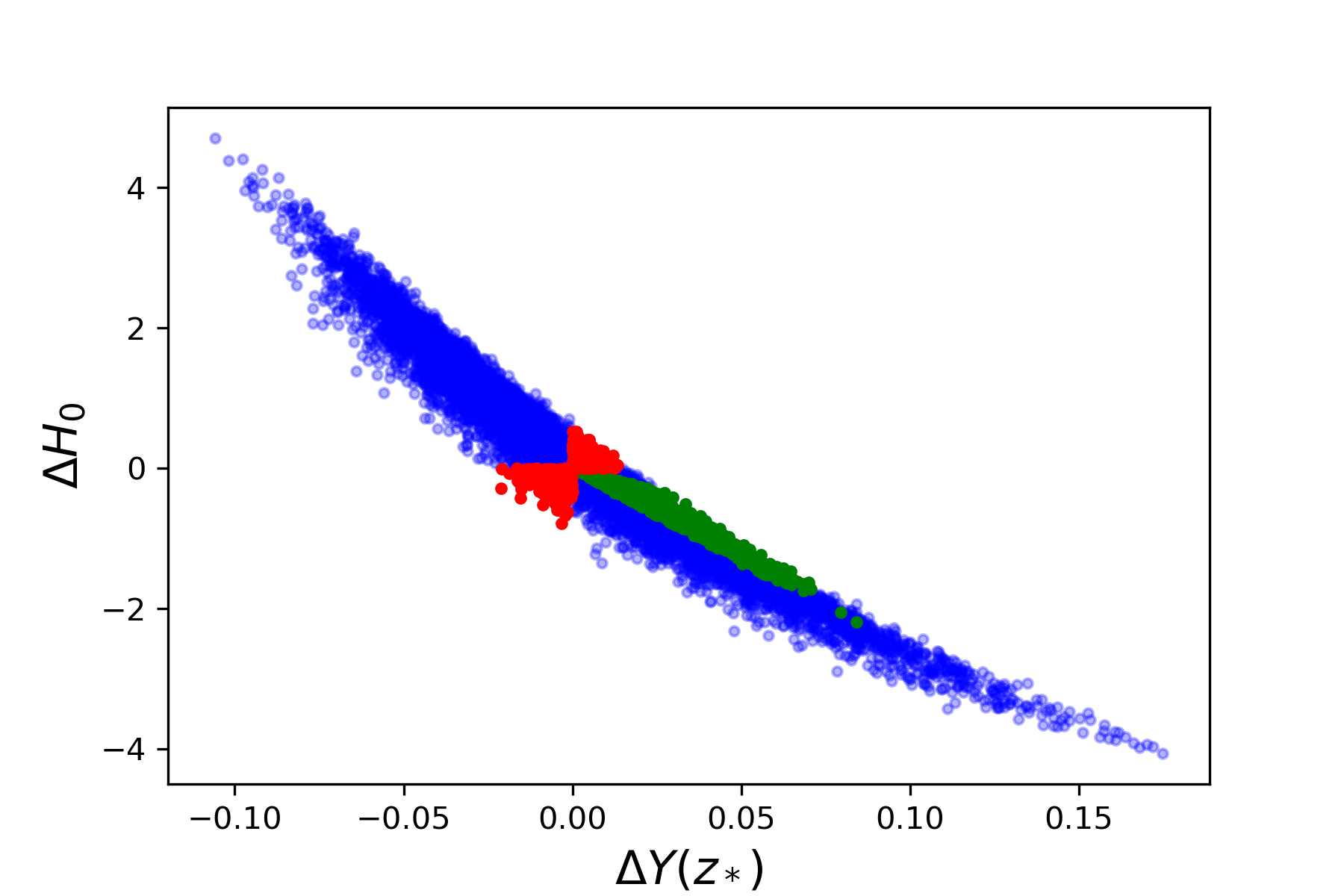}
\caption{Differences in $H_0$ in km/s/Mpc versus integrated DE density $Y(z_{*})$  for the CPL \cite{Chevallier:2000qy, Linder:2002et} and flat $\Lambda$CDM model. Each dot represents best fits to a mock realisation of data consistent with Planck-$\Lambda$CDM, namely forecasted DESI BAO in the redshift range $0 < z \leq 3.55$ and a high redshift CMB prior on $\Omega_{m} h^2$. Blue points are consistent with  anti-correlated $H_0$ and $Y(z_*)$, red points represent violations and green dots represent CPL models with $w_{\textrm{DE}}(z) > -1$ for all $z$.}
\label{fig:CPL_H0Y}
\end{figure}

\paragraph{Mock data.} Having explained why $X(z)$, or its integrated quantity $Y(z)$, and $H_0$ should be correlated, let us turn to the mock data, so that we can substantiate the correlation. In performing fits to mock data, we make a number of assumptions. First, we assume observational data consistent with $\Lambda$CDM, more accurately Planck-$\Lambda$CDM, drawn from a DESI BAO forecast \cite{DESI:2016fyo} and a Gaussian CMB prior on the combination $\Omega_{m} h^2 = 0.141 \pm 0.006, h:=H_0/100$ (see appendix of \cite{Krishnan:2021dyb}). For our purposes the former simply serves as a basis to construct mock Hubble parameter $H(z)$ and angular diameter distance $D_{A}(z)$ data in the redshift range $0 < z \leq 3.55$, while the latter represents a generous prior that comes from removing the low $\ell < 30$ multiples from CMB data \cite{Vonlanthen:2010cd, Audren:2012wb, Audren:2013nwa, Verde:2016wmz}. Note that relative to Planck-$\Lambda$CDM \cite{Planck:2018vyg}, this inflates the error considerably, thus allowing us to reduce the sensitivity of CMB data to the specifics of the DE model \cite{Vonlanthen:2010cd, Audren:2012wb, Audren:2013nwa, Verde:2016wmz}. The key point here is that we have some input from CMB at high redshift and some representative low redshift data, both of which are consistent with Planck-$\Lambda$CDM. Since we are using BAO forecasts, the reader may complain that it is overly presumptive to assume future DESI data releases will agree with Planck-$\Lambda$CDM. In light of the fact that Lyman-$\alpha$ BAO is already discrepant with Planck-$\Lambda$CDM \cite{BOSS:2014hwf, duMasdesBourboux:2020pck}, this is true. However, we can address this point by later limiting the redshift range below $z=1$, where neglecting a recent DES result \cite{DES:2021esc}, findings are largely consistent with Planck-$\Lambda$CDM. Finally, observe that since we are mainly interested in relative differences in $H_0$, one is free to shift $H_0$ up and down in the mocking procedure and the conclusions will not change. 

On a related note, mocks ultimately teach us very little about absolute differences in parameters, e. g. $H_0$, since it is the assumptions in the mocking, namely the mean value of the Hubble parameter and the quality of the data that determine displacements. In other words, the errors are put in by hand. Thus, in our work, absolute displacements in $H_0$ or $\Delta H_0$ are not meaningful and they should not be compared to existing discrepancies in \textit{real} data. Instead, once one uses a fixed mocking procedure throughout, one can look for general trends and make comparisons between cosmological models. Thus, we will not be able to say if a DE model can fully resolve $H_0$ tension or not, but based on our mocks, we can identify the models that perform better.

In Fig. \ref{fig:CPL_H0Y} we show the correlations between $\Delta Y(z_{*})$ and $\Delta H_0 := H_0^{\textrm{CPL}}-H_0^{\Lambda\textrm{CDM}}$ for approximately $10,000$ mock realisations of data based on the flat $\Lambda$CDM model.\footnote{In a number of mocks we find a deceleration-acceleration transition followed by a later acceleration-deceleration transition and (\ref{func}) has multiple roots. Removing such possibilities reduces the number of mocks below $10,000$.}  Concretely, we adopt forecasted DESI redshifts and errors for $H(z)$ and $D_{A}(z)$ assuming the redshift range $0 < z \leq 3.55$ and sky coverage of 14,000 deg$^2$ \cite{DESI:2016fyo}, which we reproduce in appendix \ref{sec:DESI}. We also adopt the CPL parametrisation, $w(z) = w_0  + w_a z/(1+z)$ \cite{Chevallier:2000qy, Linder:2002et}, and mock up on the Planck-$\Lambda$CDM values \cite{Planck:2018vyg}, $H_0 = 67.36, \Omega_{m} = 0.3153, w_0 = -1, w_a = 0$, while allowing the Gaussian prior $\Omega_{m} h^2 = 0.141 \pm 0.006$. In Fig. \ref{fig:CPL_H0Y} each point or dot represents the difference in $H_0$ and $Y(z_*)$ between CPL and flat $\Lambda$CDM for a different mock realisation. In blue we record mocks where $\Delta H_0$ and $\Delta Y(z_*)$ are anti-correlated. The green dots represent a subset of the blue dots where we demand $w_0 > -1$ and $w_0 + w_a > -1$ \cite{Vagnozzi:2018jhn}, thereby ensuring that $w_{\textrm{DE}}(z) > -1$ for all redshifts. Unfortunately, the anti-correlation between $\Delta H_0$ and $\Delta Y(z_*)$ is not a strict one and the red dots correspond to a few hundred mocks where increases (decreases) in $Y(z_*)$ are correlated with increases (decreases) in $H_0$. These exceptions ultimately undermine the utility of integrated DE density on its own as a diagnostic for higher or lower values of $H_0$. Nonetheless, the above analysis already provides good intuition.

\begin{figure}[htb]
  \centering
\includegraphics[width=90mm]{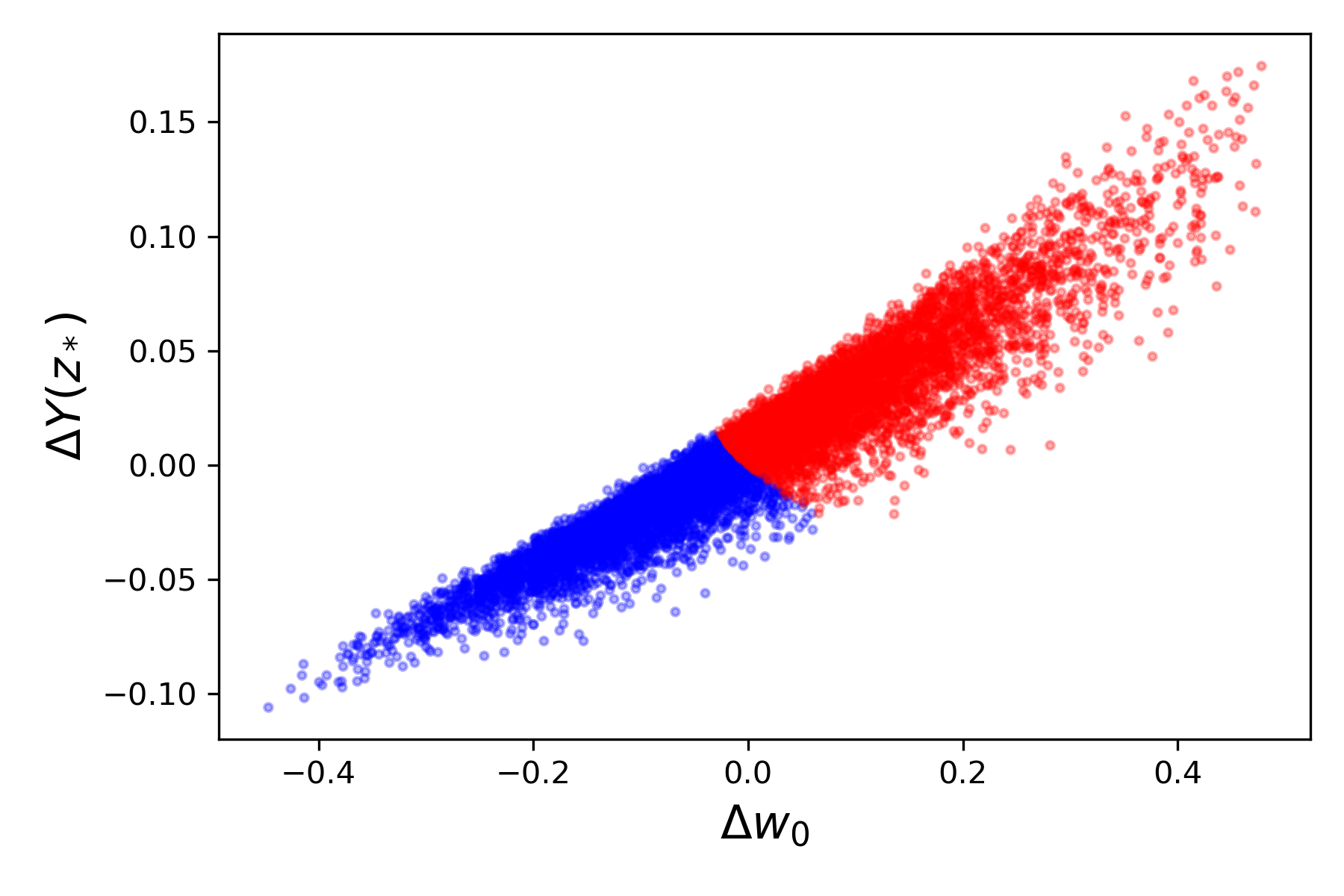}
\caption{The output of best-fits of the CPL model and flat $\Lambda$CDM to $\sim 10,000$ mock realisations of BAO data. Blue dots and red dots denote mocks where $H_0$ increases and decreases, respectively, relative to flat $\Lambda$CDM.}
\label{fig:CPL_Yw0}
\end{figure}

However, from an analysis of the remaining red dots, one identifies an anti-correlation between $w_0$ and $H_0$. More concretely, for the red dots where $H_0$ increases, it is always the case that $w_0 < -1$, and \textit{vice versa}. In Fig. \ref{fig:CPL_Yw0}, we plot the same mocks but in the $\Delta Y(z_*)$ and $\Delta w_0 \equiv w_0 + 1$ plane, where blue dots and red dots distinguish mocks where $H_0$ increases and decreases, respectively. {As we have just noted, if $\Delta H_0 > 0$, one can separate the (blue) points into $\Delta Y(z_*) <0$ and $\Delta w_0 < 0$. That being said, what one would like is simply the opposite, namely to infer $\Delta H_0$ from $w_{\textrm{DE}}(z)$. It should be clear that neither $\Delta w_0 < 0$ nor $\Delta Y(z_*) < 0$ is strong enough on its own to leave one with only points where $H_0$ increases, but if both $\Delta w_0 <0$ and $\Delta Y(z_*) < 0$, then $H_0$ is guaranteed to increase. On the flip side, if $\Delta w_0 >0$ and $\Delta Y(z_*) > 0$, then $H_0$ must decrease. As is clear from Fig. \ref{fig:CPL_Yw0}, the criteria do not cover all the mocks, but it is easy to check that mocks where $\Delta w_0$ and $\Delta Y(z_*)$ have different signs, i. e. $\Delta w_0 \cdot \Delta Y(z_*) < 0$, any displacements in $H_0$ are well within $1 \sigma$. We have marginalised over the parameters ($H_0, \Omega_m, w_0, w_a)$ for a number of the mocks using Markov Chain Monte Carlo (MCMC) and found that the $1 \sigma$ confidence interval for $H_0$ is $\sigma_{H_0} \gtrsim 1.3 $ km/s/Mpc. In contrast for mocks where $\Delta w_0 \cdot \Delta Y(z_*) <0$ in Fig. \ref{fig:CPL_Yw0}, we find the maximum and minimum values of $\Delta H_0$ are $\Delta H_0 = 0.56$ and $\Delta H_0 = -0.78$, respectively. As stated, these displacements are well within representative $H_0$ errors. The key point is that if one wants to have an increase or decrease in $H_0$, then $\Delta w_0$ and $\Delta Y(z_*)$ must possess the same sign. 

\begin{figure}[htb]
  \centering
  \begin{tabular}{cc}
\includegraphics[width=75mm]{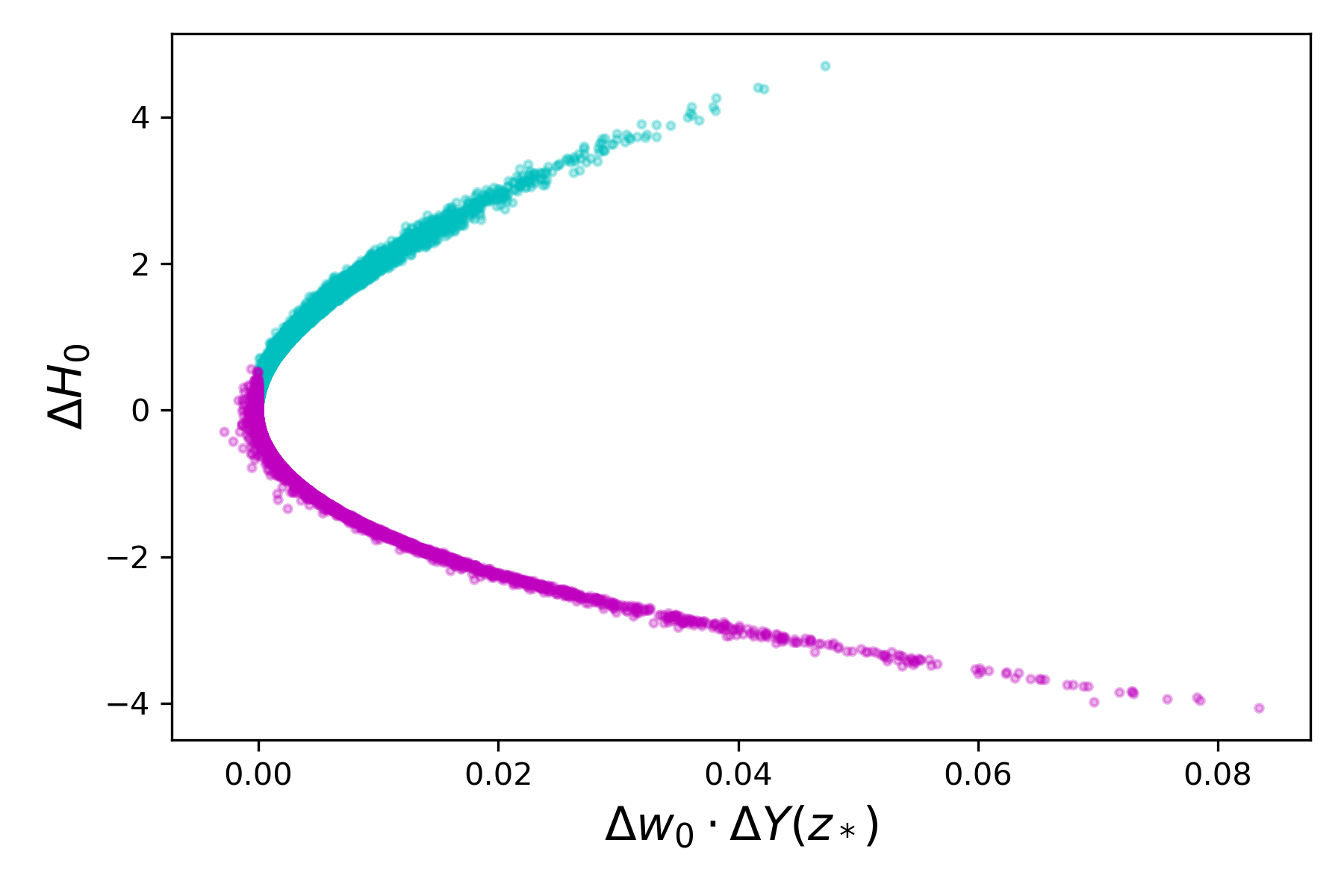}&
\includegraphics[width=75mm]{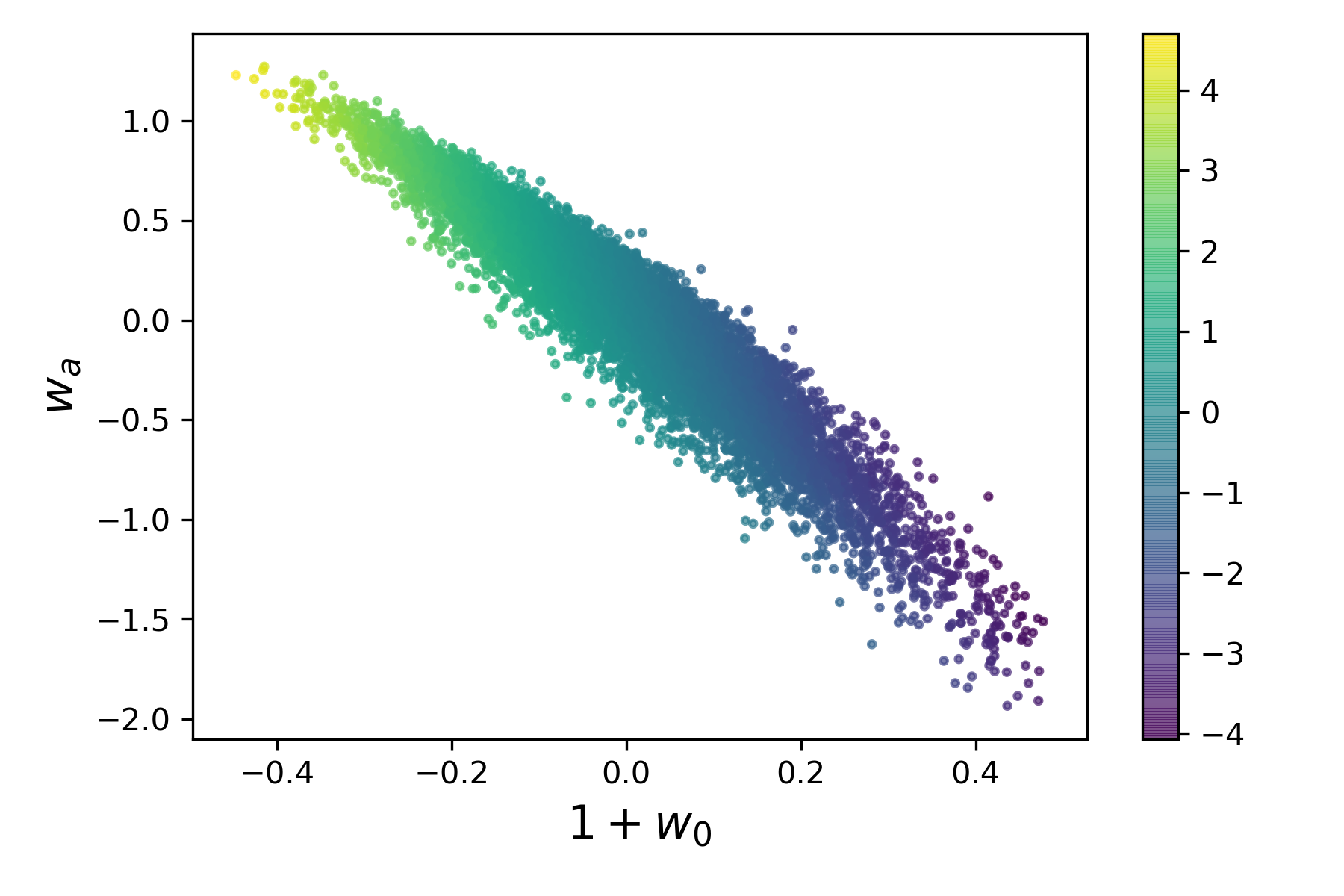}
\end{tabular}
\caption{Left plot: we distinguish mocks in cyan where \textit{both} $\Delta w_0 < 0$ and $\Delta Y(z_*) < 0$ from the remaining mocks in magenta. Evidently, unless both conditions are satisfied, any increase in inferred $H_0$ is marginal at best. Right plot: we show differences in $H_0$ values as given by the color-coding of the legend in the right of the plot, in the more familiar $(w_0, w_a)$-plane. Once again, larger increases in $H_0$ (yellow) are driven by $w_0$, whereas $w_a$ simply compensates, since mock data are consistent with flat $\Lambda$CDM. Note that the difference between the SH0ES $H_0$ determination \cite{Riess:2021jrx} and Planck \cite{Planck:2018vyg} is about $5.5$ km/s/Mpc.}
\label{fig:w0wa}
\end{figure}

Moving along, in Fig. \ref{fig:w0wa}, Left plot, we show CPL models that increase $H_0$ the most not only have $\Delta w_0 < 0$ but also $\Delta Y(z_*) < 0$.  In the Right plot of Fig. \ref{fig:w0wa} we show the same mocks in the more familiar $1+w_0$ and $w_a$ plane to demonstrate that significant increases in $H_0$ (yellow) are driven by $w_0$. There is one other take home message from the plot. One can clearly see that the data is consistent with flat $\Lambda$CDM, since fits with $w_0 < -1$ are correlated with $w_a > 0$. Note that if $w_a > 0$, this marks an increasing trend in $w_{\textrm{DE}}(z)$. In short, one cannot deviate far from the $\Lambda$CDM EoS, $w_{\textrm{DE}}(z) = -1$. We will come back to this point later when we study EFTs in the Horndeski class. Let us simply remark that for this reason, coupled Quintessence models \cite{Amendola:1999er} are less effective at alleviating $H_0$ tension precisely because  $w_{\textrm{DE}}(z=0) > -1$ \cite{Das:2005yj}. More precisely, the value of the coupling at $z=0$ is degenerate with matter density $\Omega_{m}$, so one can always rescale $\Omega_{m}$ to remove the coupling at $z=0$, in which case one is in the $w_{\textrm{DE}} > -1$ regime close to $z=0$ \cite{Das:2005yj}. {As a result, one expects any increase in $H_0$ within a coupled Quintessence model to be within $1 \sigma$. Note, we have not imposed a local $H_0$ prior and the $H_0$ inferences we make are simply driven by cosmological {mock} data.} 

Obviously, our findings rest exclusively on the CPL model over an extended redshift range $0 < z \leq 3.55$. Given that Lyman-$\alpha$ BAO is discrepant with Planck-$\Lambda$CDM at $z \sim 2.3$ \cite{BOSS:2014hwf, duMasdesBourboux:2020pck}, it is prudent to restrict the redshift range below $z=1$ and make sure that this does not change the result. The main point is that current BAO results below $z=1$ are largely consistent with Planck-$\Lambda$CDM (however, see \cite{DES:2021esc}), whereas we should be open to deviations occuring at higher redshifts. Moreover, we should also change the DE parametrisation and document any changes, since as stressed in \cite{Colgain:2021pmf}, each $(w_0, w_a)$ parametrisation represents an arbitrary choice and one should check if the statements are robust across parametrisations, e g. \cite{Yang:2021flj}. To that end, we consider the Taylor expansion in $z$, $w_{\textrm{DE}}(z) = w_0 + w_a z$ \cite{Cooray:1999da, Astier:2000as}. This redshift model is arguably even simpler than CPL, but represents a dubious expansion beyond $z=1$, since $z$ is no longer a small parameter. The results of the exercises are shown in Fig. \ref{fig:lowz_Taylor}. Once again, we find that {the more significant displacements in $H_0$ are driven by EoS where $\Delta w_0$ and $\Delta Y(z_*)$ have the same sign. Unsurprisingly, as we restrict the data below $z=1$, we see that it is less constraining, so displacements in $H_0$ can become larger, otherwise Fig. \ref{fig:lowz_Taylor} is in line with our expectations from Fig. \ref{fig:w0wa}.}

\begin{figure}[htb]
   \centering
\begin{tabular}{cc}
\includegraphics[width=75mm]{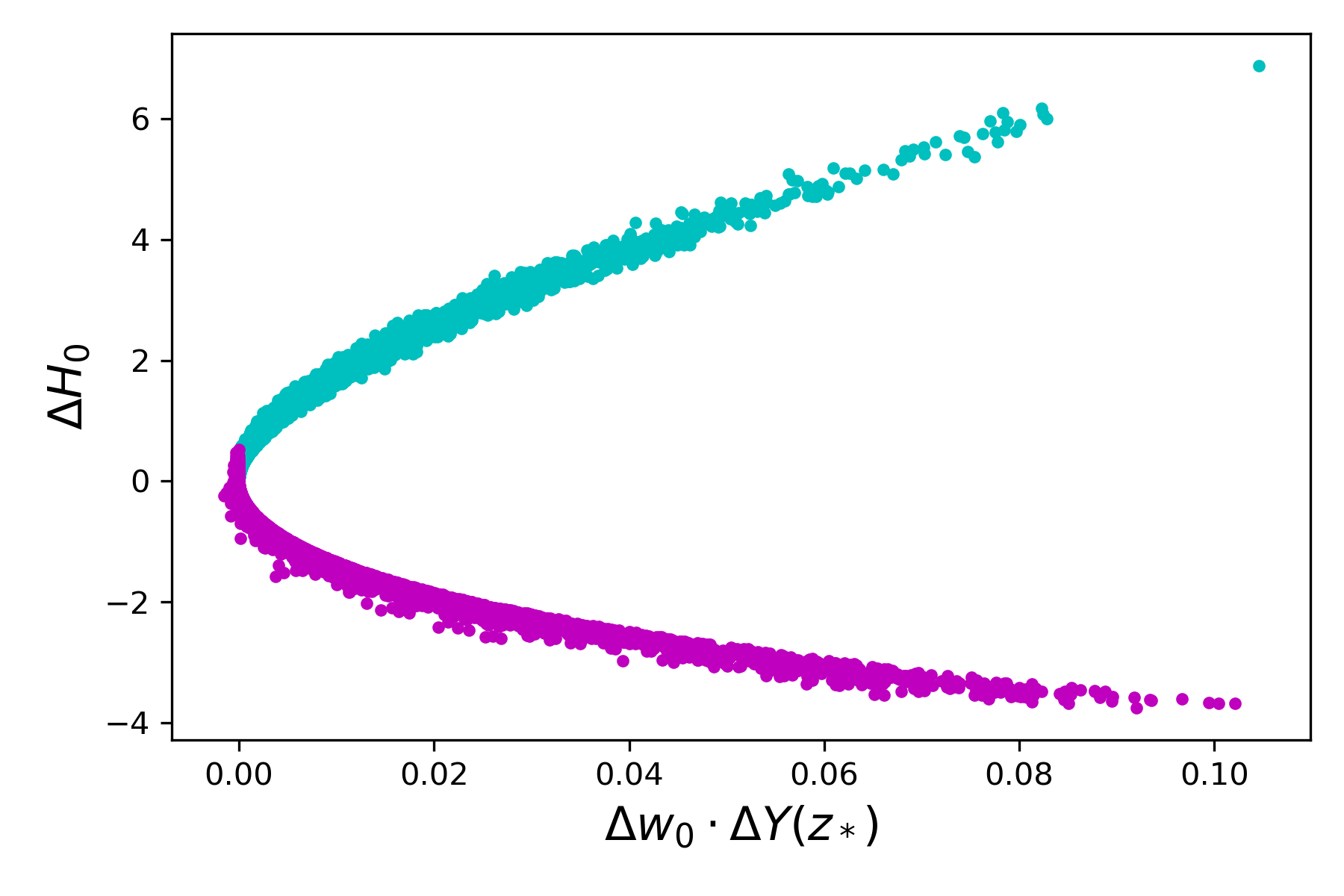}&
\includegraphics[width=75mm]{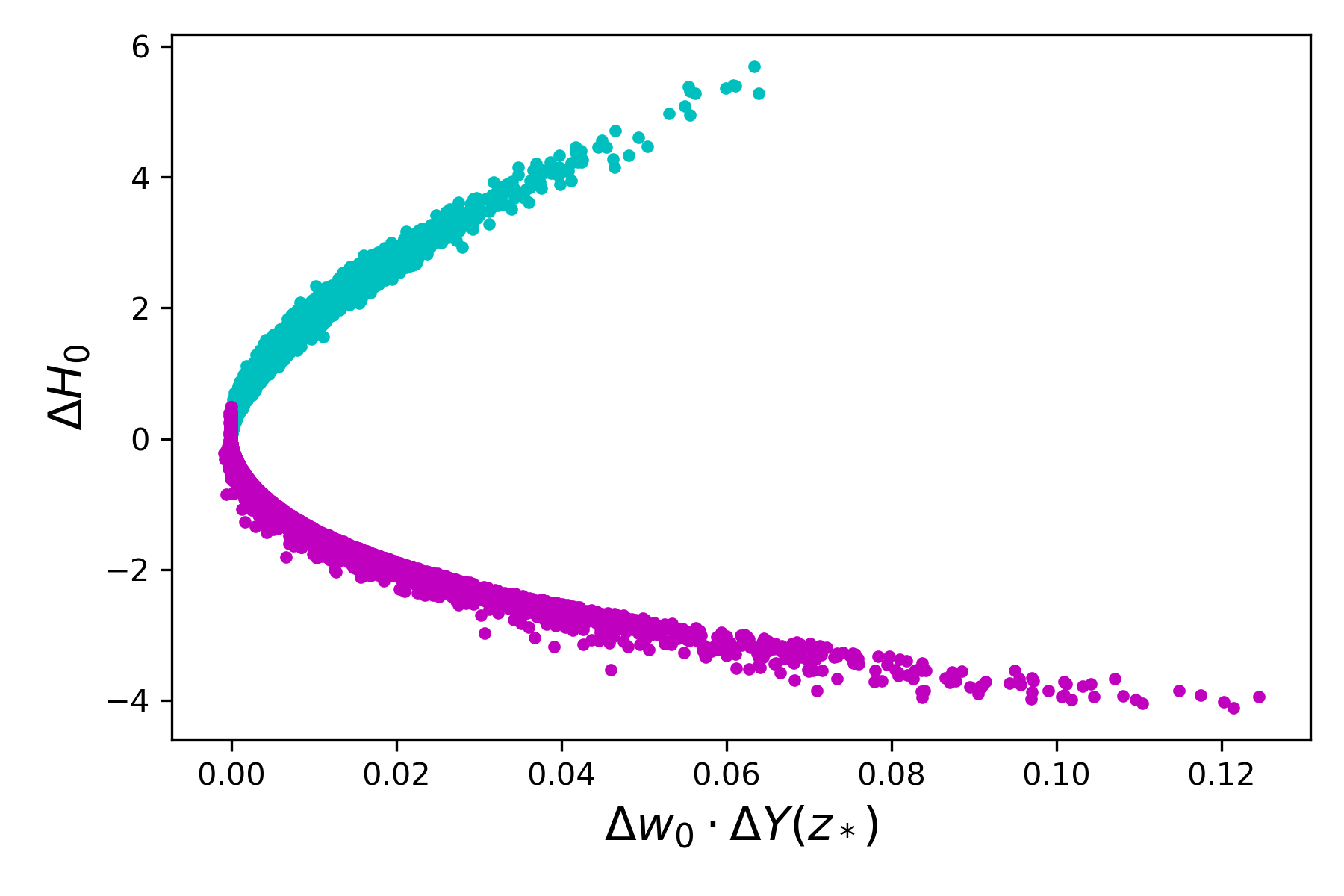}
\end{tabular}
\caption{Same as Fig. \ref{fig:w0wa}, but with the restricted redshift range $z < 1$ for mock data. On the left we present results for the CPL model, and on the right, for a Taylor expansion in redshift. Compared to Fig. \ref{fig:w0wa}, we see that  $\Delta H_0$ has understandably increased, since the data below $z=1$ is less restrictive.}
\label{fig:lowz_Taylor} 
\end{figure}

Let us summarise. Based on our analysis of $(w_0, w_a)$ parametrisations for evolving DE, we have noted for a large number of mocks consistent with flat $\Lambda$CDM that $H_0$ always increases relative to $\Lambda$ if {both} $\Delta Y(z_*) < 0$ {and} $1+w_0 < 0$. The former quantity allows one to incorporate evolution. On the flip side, if $\Delta Y(z_*) > 0$ and $1+w_0 > 0$, then $H_0$ is guaranteed to decrease. {For configurations where differences in $w_0$ and $Y(z_*)$ have the opposite sign, one can expect any displacements in $H_0$ to be marginal and within $1 \sigma$. Given that $H_0$ tension may be anything up to a $\sim 5-6 \sigma$ discrepancy, this should preclude models with $\Delta w_0 \cdot \Delta Y(z_*) < 0$.} It should be clear that there is some redundancy in $\Delta w_0$ and $\Delta Y(z_*)$ as $w_0$ dictates the behaviour of  $w_{\textrm{DE}}(z)$ in a window around $z=0$, however the integrated quantity, $Y(z_*)$, is important when one has evolution. It is now easy to infer that $H_0$ must decrease if one replaces $\Lambda$  with a K-essence model \cite{Armendariz-Picon:1999hyi, Chiba:1999ka, Armendariz-Picon:2000ulo}. To appreciate this, recall that the DE EoS is 
\be
\label{kessence_eos}
1+w_{\textrm{DE}} =  \frac{2XG_{2, X}}{2X G_{2, X}-G_2}. 
\ee
The absence of ghost and instabilities of scalar perturbations implies $G_{2,X}> 0$, $(\rho_{\text{DE}})_{,X}=2XG_{2,XX}+G_{2,X}>0$ where $\rho_{\textrm{DE}} = 2X G_{2, X}-G_2$ is the DE  density (see discussion in  \cite{Kase:2018aps}). Assuming the weak energy condition  $\rho_{\textrm{DE}}\geq 0$, we have $w_{\textrm{DE}}(z) > -1$, implying that $H_0$ must decrease when the model is fitted to any data consistent with flat $\Lambda$CDM. This has already been explicitly established for the Quintessence class by analysing the dynamics of the field theory \cite{Banerjee:2020xcn}. {A secondary lesson is that models where $\Delta w_0 \cdot \Delta Y(z_*) < 0$ can only marginally alleviate $H_0$ tension. The reason being that one can always absorb a coupling in the matter density at $z=0$, so that $w_{\textrm{DE}} > -1$ near $z=0$. This highlights an obvious $H_0$ problem for coupled Quintessence \cite{Amendola:1999er, Das:2005yj}}. 

\section{Non-minimal coupling}
\label{sec:nonmin}
Having warmed up in the last section by discussing Quintessence and K-essence, we turn our attention to the next simplest Horndeski models by permitting a non-minimal coupling to gravity. Concretely, we restrict our attention to the Quintessence sector, $G_2(\phi, X) = X- V(\phi)$, redefine $F({\phi}) = 2 G_{4} (\phi)$, which allows us to connect to other works in the literature \cite{Boisseau:2000pr, Esposito-Farese:2000pbo}, and set $M_{\textrm{pl}} =1$ ($G_{4} = \frac{1}{2})$. Throughout we will be assuming that $F(\phi)$ is an analytic function, otherwise one may have to {physically explain an abrupt jump in $G$}, e. g. \cite{Marra:2021fvf, Alestas:2021luu, Alestas:2022xxm, Perivolaropoulos:2022vql}.\footnote{The mocking procedure in this work makes use of BAO. It is fitting to note that variations in Newton's constant have interesting implications for Type Ia supernovae as standard candles \cite{Amendola:1999vu, Garcia-Berro:1999cwy}, but this does not concern us.} See  \cite{Odintsov:2022eqm} for a  potential explanation in $f(R)$ gravity. It is worth stressing that any evolution in $F(\phi)$ translates into an evolution in Newton's constant $G$. Currently, observations are largely consistent with no evolution, whether it be constraints in the solar system from lunar laser ranging $\dot{G}/G_0 =  (7.1 \pm 7.6) \times 10^{-14}$ yr$^{-1}$ \cite{Hofmann:2018myc}, or constraints from the Big Bang nucleosynthesis (BBN), $G_{\textrm{BBN}}/G_0 = 0.98 \pm 0.06$ \cite{Alvey:2019ctk}, where $G_0$ denotes the value of Newton's constant today and the quoted errors in the latter are at the $2 \sigma$ level. Nevertheless, within these constraints, one is always free to speculate.  

Our goal here is to recycle the analysis in \cite{Banerjee:2020xcn}, which allowed us to treat Quintessence models in a model agnostic fashion through Taylor expansions in the late universe. There, expanding the scalar $\phi$ in redshift $z$ - a ``small'' parameter below $z = 1$ - about its value today $\phi_0$, one can also expand the Quintessence potential provided the displacement in the scalar is also small, $|\phi - \phi_0| < 1$. Imposing the Quintessence equations of motion order by order in $z$, the parameters in the potential are fixed in terms of scalar parameters, the Hubble constant $H_0$ and matter density $\Omega_{m}$. This is the basic picture, however, here we have an extra scalar function, $F(\phi)$, in addition to the potential $V(\phi)$, which leads to additional unconstrained parameters when expanded in $z$. That being said, as explained above, we have a tight constraint from BBN \cite{Alvey:2019ctk}, which once translated into $F$, becomes  
\be
\label{BBN}
F_{\textrm{BBN}}/{F_0} = 1.02 \pm 0.06, 
\ee
where $F_0$ is the value of $F(\phi)$ today and once again the constraint is at the $2 \sigma$ level.  Here, it is interesting to note that one can analyse CMB data while allowing $G$ to vary, but apart from inflating errors, it does not greatly affect the central values \cite{Bai:2015vca}, so the prior can be justified. As discussed, we need to make some assumption about the evolution in $F(\phi)$, but since there is no compelling evidence for evolution, we follow \cite{Alvey:2019ctk} and make the simplest assumption that $F$ evolves linearly in time. While instabilities can arise if $F < 0$ \cite{Matsumoto:2017qil}, the strong constraint (\ref{BBN}) coupled with linear evolution safely precludes this possibility. Once this is done, one can immediately write down an expression for $F(z)$: 
\be
\label{lookback}
F(z) = F_0 + \dot{F} \, \tau(z), \quad \tau(z) =  \frac{977.8}{H_0} \int_0^z \frac{ \dd z^{\prime}}{(1+z^{\prime}) E(z^{\prime})} \textrm{ Gyr}, 
\ee
where $E(z)=H(z)/H_0$ is the normalised Hubble parameter. In addition, we have introduced the look-back time $\tau(z)$ in units of Gyr \cite{Bernal:2021yli} and $\dot{F}$ is simply the slope or derivative of $F$ with respect to look-back time. Noting that $\tau (z = \infty) \approx 13.8$ Gyr in the Planck-$\Lambda$CDM Universe \cite{Planck:2018vyg}, this allows us to infer a constraint on the slope so that $F( z_{\textrm{BBN}}) = F_{\textrm{BBN}}$, 
\be
\label{normalF}
F_1 := 977.8 \dot{F} =  ( 0.02 \pm 0.06) H_0.
\ee
As one can see, the rescaled slope $F_1$ is less than 8\% of the value of $H_0$ at $2 \sigma$. Evidently, when compared to $H_0$ this is not negligible, and as we will see, even linear temporal evolution in $F(\phi)$ can alter the conclusions from the previous section. The reader will note that we started with an additional function $F(\phi)$, but through the above linear assumption, it is reduced to a single constant parameter $F_1$. As a result, any change to our perturbative analysis within Quintessence following  \cite{Banerjee:2020xcn} is minimal, since we have only introduced an additional constant.   

Returning to the perturbative, low redshift analysis, if we expand the Hubble parameter to cubic order in redshift, 
\be
\label{expandedH}
H(z) = H_0 (1 + h_1 z + h_2 z^2 + h_3 z^3 + \dots), 
\ee
it is easy to write down a perturbative expression for $F(z)$ that is expected to be valid at redshifts $z \lesssim1$: 
\bea
\label{F}
F(z) = 1 + \frac{F_1}{H_0} \biggl(z &-& \frac{1}{2} (1+ h_1) z^2 + \frac{1}{3} (1 + h_1 + h_1^2-h_2) z^3 \nn &-& \frac{1}{4} ( 1+ h_1^2+h_1^3-h_2 + h_1-2 h_1 h_2 + h_3) z^4  + \dots \biggr), 
\eea
where we have set $F_0 = 1$ without any loss of generality, since one can always absorb a constant contribution to $F(z)$ in the Newton constant today $G_0$. This expression effectively determines $F(z)$ in terms of the expanded Hubble parameter $H(z)$ and the slope $F_1$, subject to the requirement that the latter satisfies (\ref{normalF}) at $2 \sigma$. Since we have expanded the Hubble parameter, it should be noted that our expression for $F(z)$ is an approximation. Nevertheless,  any error introduced by this approximation is expected to be negligible, because  despite the approximation we have used for the integral, only being accurate to $20 \%$ through to $z=1$ for $\Omega_{m} \approx 0.3$, relative to the leading term in (\ref{F}), any discrepancy is small. Indeed, one can check that the error inherent in the approximation for the Planck-$\Lambda$CDM cosmology, where $h_1 \approx 0.45, h_2 \approx 0.35, h_3 \approx -0.007$, is less than $1 \%$ through to $z = 1$. It should also be stressed that by imposing the Gaussian constraint (\ref{normalF}) on $F_1$ we can access variations in $F(z)$ where $F$ increases and decreases with redshift, of course consistent with the BBN constraints \cite{Alvey:2019ctk}. It is worth noting that since the central value of $F_{BBN} > 1$, our setup has a preference for increasing values of $F(z)$, or alternatively decreasing values of $F(t)$, where $t$ is cosmological comoving time. 

Having set the problem up, we can now turn our attention to the equations \cite{Boisseau:2000pr, Esposito-Farese:2000pbo}, 
\begin{subequations}\label{EOM}
\begin{align}
3 F H^2 =& \rho_m + \frac{1}{2} \dot{\phi}^2 + V - 3 H \dot{F}, \label{EOM1}\\
-2 F \dot{H} =& \rho_m + \dot{\phi}^2 + \ddot{F}-H \dot{F}, \label{EOM2}\\
0 =& \ddot{\phi} + 3 H \dot{\phi} + \partial_{\phi} V - 3 ( \dot{H} + 2 H^2) \partial_{\phi} F,  \label{EOM3}   
\end{align}
\end{subequations}
where \textit{dot} denotes derivative w.r.t the comoving time $t$. Recalling that $\dot{\rho}_{m} + 3 H \rho_m = 0$, only two of the three equations are independent. We can now make a simple hand waving observation relating our analysis to  \cite{Heisenberg:2022lob, Heisenberg:2022gqk},  where it is  found that variations in Newton's constant $G$ alleviate $H_0$ tension provided $\delta G < 0$. For slowly evolving $F \propto G^{-1}$, one can rewrite  (\ref{EOM1}) in the same form as (\ref{hubble}), where it should be noted that increasing $F$ (decreasing $G$) leads to a decrease in the DE density $X(z)$. Therefore, just at the level of the equations of motion, the conclusions in \cite{Heisenberg:2022lob, Heisenberg:2022gqk} are in line with expectations given the above equations of motion. A larger local $H_0$ necessitates $\dot{F} > 0$. 

Our choice for linear evolution of $F$ with time means that $\ddot{F} = 0$ and that $\dot{F}$ follows from the choice of the constant $F_1$, $\dot{F} = -F_1$.  A difference in time can be attributed to the fact that when one solves equations one is implicitly integrating forwards in time, whereas $\tau$ (\ref{lookback}) denotes the reverse or look-back time. We can now replace derivates $\frac{\dd}{\dd t} \rightarrow - H (1+z) \frac{\dd}{\dd z} $ and rewrite $\rho_{m} = 3 H_0^2 \Omega_{m} (1+z)^3$ to bring the equations to the simpler form: 
\be
\begin{split}
3 F H^2 =& 3 H_0^2 \Omega_{m} (1+z)^3 + \frac{1}{2} (1+z)^2 H^2 (\phi^{\prime})^2 + V(\phi) + 3 H F_1, \\
2 F H (1+z) H^{\prime} =& 3 H_0^2 \Omega_{m} (1+z)^3 + (1+z)^2 H^2 (\phi^{\prime})^2 + H F_1,  
\end{split}
\ee
where \textit{prime} now denotes derivative w.r.t $z$ and an explicit expression for $F(z)$ was given in equation (\ref{F}). 

We can now solve these equations perturbatively at lower redshifts $z < 1$. The advantage of this approach is that one can treat the potential $V(\phi)$ in a model agnostic fashion \cite{Banerjee:2020xcn}. The basic idea is to expand $\phi(z)$ in terms of $z$ about its value today $\phi(z=0) = \phi_0$, on the assumption that $z$ is small ($z < 1$), while at the same time expanding the potential in terms of $\phi-\phi_0$:\footnote{One can perform the expansion in powers of $\Delta\phi:=\phi-\phi_0$, instead of powers of $z$. Physically this corresponds to taking $\phi$ as the cosmic clock.}
\be\label{phiV}
\begin{split}
\phi =& \phi_0 + \alpha z + \beta z^2 + \gamma z^3 + \dots , \\
V(\phi) =& V_0 + V_1 (\phi - \phi_0) + V_2 (\phi - \phi_0)^2 +  \dots, 
\end{split}
\ee
where the triplet of constants ($\alpha, \beta, \gamma)$ allow the scalar to be dynamical and the $V(\phi)$ expansion is valid provided $|\phi - \phi_0| < 1$. Solving order by order in $z$, we find
\be\label{potential}
\begin{split}
\frac{V_0}{H_0^{2}} =& 3 (1 - \Omega_{m}) - \frac{1}{2} \alpha^2 - 3 \frac{F_1}{H_0}, \\
\frac{V_1}{H_0^{2}} =& - \frac{1}{2} \alpha^3 - 2 \beta + 2 \alpha - \frac{3}{2} \alpha \Omega_m + \frac{1}{2 \alpha} \frac{F_1}{H_0} (12 - 4 \alpha^2 - 9 \Omega_{m}) - \frac{3}{2 \alpha}  \frac{F_1^2}{H_0^2} , \\
\frac{V_2}{H_0^{2}} =& - \frac{1}{4} \alpha^4 + \frac{1}{4} \alpha^2 (1-3 \Omega_m)- \frac{5}{2} \alpha \beta + 1  - \frac{3}{2 \alpha} (3 \beta  \Omega_m +2 \gamma ) \\
+& \left( - \frac{1}{2} \alpha^2 - 3 \frac{\beta}{\alpha} + \frac{3}{8} ( 4 + \Omega_{m}) - \frac{3 \beta}{2 \alpha^3} (4 - 3 \Omega_{m}) - \frac{3}{8 \alpha^2} (8 - 9 \Omega_{m}^2) \right) \frac{F_1}{H_0}  \\ 
+& \left( \frac{9}{4 \alpha^2} + \frac{1}{2}  + \frac{3 \beta}{2 \alpha^3}  + \frac{27}{8 \alpha^2} \Omega_{m} \right) \frac{F_1^2}{H_0^2}  + \frac{3}{4 \alpha^2} \frac{F_1^3}{H_0^3}, 
\end{split}
\ee
and 
\be\label{hi}
\begin{split}
h_1 =& \frac{1}{2} \alpha^2 + \frac{3}{2} \Omega_{m} + \frac{1}{2} \frac{F_1}{ H_0}, \\
h_2 =&  \frac{1}{8} \alpha^4 + \frac{1}{4} \alpha^2 + \alpha \beta + \frac{3}{8} \Omega_m (4 - 3 \Omega_m) - \frac{1}{8} \frac{F_1}{H_0} (2 + \alpha^2 + 9 \Omega_{m}) - \frac{1}{4} \frac{F_1^2}{H_0^2}, \\
h_3 =& \frac{\alpha^6}{48} + \frac{\alpha^4}{16} (\Omega_{m} + 2) + \alpha \gamma  + \frac{\alpha^2}{16} \Omega_m ( 9 \Omega_{m} -2) + \frac{\alpha \beta}{2} ( \alpha^2 + \Omega_{m} + \frac{4}{3}) + \frac{2}{3} \beta^2 
\\ +& \frac{1}{16} \Omega_{m} (8 - 36 \Omega_{m}+27 \Omega_{m}^2) + \frac{1}{48} \big(8 - 3 \alpha^4- 16 \alpha \beta -54 \Omega_{m} \\ +& 117 \Omega_{m}^2 + 2 \alpha^2 (9 \Omega_{m} -1) \big) \frac{F_1}{H_0}  
+ \frac{1}{8} \left( 2 + \alpha^2 + 10 \Omega_{m} \right) \frac{F_1^2}{H_0^2} + \frac{5}{24} \frac{F_1^3}{H_0^3}. 
\end{split}
\ee
As a consistency check, one can check that when $F_1 = 0$ we recover the expressions in  Ref. \cite{Banerjee:2020xcn}. Observe also that without the coupling, $F_1=0$, the potential is fixed by the equations of motion in terms of $(\alpha, \beta, \gamma)$ and the observational parameters, $(H_0, \Omega_{m})$. In solving these equations, we made sure to solve to one order higher by also incorporating expressions for $V_3$ and $h_4$, which are lengthy, so we have omitted them. In the process of solving perturbatively, one can eliminate the constant parameters describing the potential in terms of the constant parameters in the scalar. Thus, the free parameters are $H_0, \Omega_{m}$ from the original flat $\Lambda$CDM model and an additional $\alpha, \beta, \gamma$ describing the DE sector. In addition, we would have many other parameters associated to the coupling function $F(\phi)$, but we have judiciously reduced this to a single parameter $F_1$, which simply allows for linear evolution. 

Indeed, the constraints on $F_1$ are such that $F_1/H_0 \lesssim 0.1$ beyond $ 2 \sigma$. This means that higher powers of $F_1/H_0$ are going to be small, and if one wants, this allows one to simplify the expansion of the Hubble parameter. However, this does not greatly simplify expressions, so we work with the original expressions for $h_i$ above. However, even at this stage, one point should be clear. Since $F_1$ enters with opposite signs in $h_1$ and $h_2$, it is obvious from the analytic expressions that if $F_1$ lowers the slope of $H(z)$ at $O(z)$, then it will inevitably increase the slope at $O(z^2)$. This means that any flattening of the Hubble parameter in the immediate vicinity of $z \approx 0$ tends to steepen it at larger $z$. We can also see this in the DE EoS. Within our assumptions, namely linear evolution of $F$ with time, implying constant $\dot{F}$ and $\ddot{F} = 0$, the EoS of DE becomes, 
\bea
\label{w_non_min}
w_{\textrm{DE}}(z) &=& 
- 1 + \frac{(\alpha^2 + F_1 H_0^{-1}) } {3 \Omega_{\phi 0}} + \frac{z}{\Omega_{\phi 0}^2}  \biggl[ \frac{\alpha^4}{3} (\Omega_{\phi 0} -1) + \frac{\alpha^2}{3} \Omega_{\phi 0} (5 - 3 \Omega_{\phi 0} ) + \frac{4}{3} \alpha \beta \Omega_{\phi 0} \nn 
&-&  \left( \frac{\alpha^2}{6} ( 4 - \Omega_{\phi 0}) +\frac{1}{2} \Omega_{\phi 0} (1- \Omega_{\phi 0})\right) \frac{F_1}{H_0}  - \frac{(2 + \Omega_{\phi 0})}{6} \frac{F_1^2}{H_0^2} \biggr] + O(z^2)
\eea
where following  \cite{Banerjee:2020xcn} we have introduced $\Omega_{\phi 0} := 1 - \Omega_{m}$ and we have expanded in $z$. From the expressions, it is once again obvious that if $F_1$ has the right (negative) sign to decrease $w_{\textrm{DE}}(z)$ at low $z$, then it must increase $w_{\textrm{DE}}(z)$ at higher $z$ since $\Omega_{\phi 0} < 1$. In essence, any linear evolution of $F(\phi)$ with time introduces competing signs at leading and subleading orders. Even at this stage we can see that a non-minimal coupling may not work as well as the CPL model in the sense that it will be difficult to orchestrate both $\Delta w_0 < 0$ and $ \Delta Y(z_*) < 0$ in order to maximise an increase in $H_0$ (see Fig. \ref{fig:w0wa}). We will comment further on this later. We can determine the preferred sign for $F_1$ by resorting to mock data fits. 

Before getting into the mock data analysis, it is timely to review our setup as it differs from the data fitting of exact models in section \ref{sec:warmup}. Here, any given $(\alpha, \beta, \gamma)$ define a valid perturbative scalar profile provided they satisfy the following conditions \cite{Banerjee:2020xcn}, 
\be\label{constraints}
\begin{split}
| \alpha | \gtrsim z_{\textrm{max}} | \beta |, \quad | \beta | \gtrsim z_{\textrm{max}} |\gamma|, \quad |\phi- \phi_0| \lesssim 1, \\
| V_0 |  \gtrsim | V_1 \cdot (\phi - \phi_0) |, \quad |V_1 | \gtrsim |V_2 \cdot (\phi - \phi_0) |,   
\end{split}
\ee
where we choose $z_{\textrm{max}} = 1$. In appendix \ref{sec:approx} we show that the approximation is under control through to $z=1$. It was noted in \cite{Banerjee:2020xcn} that any finite $(\alpha, \beta, \gamma)$ leads to lower $H_0$ values than its flat $\Lambda$CDM counterpart, $(\alpha, \beta, \gamma) = (0, 0, 0)$. This claim can be substantiated by probing the parameter space through randomly generated  large number of triples $(\alpha, \beta, \gamma)$ in a normal distribution about $(0,0,0)$ with suitable standard deviations \footnote{Concretely, we adopt the choice $(\sigma_{\alpha}, \sigma_{\beta}, \sigma_{\gamma}) = (0.18, 0.12, 0.06)$, where we have staggered so that the first line of (\ref{constraints}) is more easily satisfied by the generated configurations.}, throwing away configurations that violate (\ref{constraints}), while for the remaining ($\alpha, \beta, \gamma$) fitting the Hubble parameter defined by (\ref{expandedH}) and (\ref{hi}) to mock data below $z=1$ to determine the best-fit values of the cosmological parameters $(H_0, \Omega_{m})$. Throughout, we adopt the flat priors $0 < H_0 < 100$ km/s/Mpc and $0 < \Omega_{m} < 1$. In practice, one does this with each generated $(\alpha, \beta, \gamma)$ and their flat $\Lambda$CDM counterpart $(\alpha, \beta, \gamma) = (0, 0, 0)$ for each iteration of mock data and compares. Throughout, we impose the same high redshift prior as section \ref{sec:warmup}, $\Omega_{m} h^2 = 0.141 \pm 0.006$. 

While Ref. \cite{Banerjee:2020xcn} performed analysis using a \textit{single realisation} of real data, including BAO, Supernovae (SN) and cosmic chronometers, and without the CMB prior, so it was simply a low redshift result, here we provide a different realisation of BAO data for each $(\alpha, \beta, \gamma)$. Repeating the same exercise, we find that over $19,075$ mock iterations satisfying (\ref{constraints}), one encounters larger values of $H_0$, with yet a better fit to the data (lower $\chi^2$) a total of 16 times. Thus, the probability of encountering such configurations is $0.08 \%$ and any increase in $H_0$ is negligible $\Delta H_0 < 0.03$ km/s/Mpc. In other words, there appear to be some exceptional perturbative Quintessence models within our assumptions, but the chance of encountering them is low. This may be an artifact of the least squares fitting procedure, but since we use the same methodology throughout, this is not expected to change results. Interestingly, we have checked that the 16 exceptional  configurations satisfy $w_{\textrm{eff}} > -1$ from equation (10) of  \cite{Banerjee:2020xcn} for uncoupled Quintessence with $C = 0$, so these exceptions do not have an explanation in a phantom EoS. It should be stressed that the mock data is based on flat $\Lambda$CDM (see section \ref{sec:warmup}), but as we explain in appendix \ref{sec:approx}, the approximation should be under control. Finally, as explained earlier, each triple $(\alpha, \beta, \gamma)$ should be viewed as a distinct Quintessence model, since the scalar profile fixes the potential through (\ref{potential}) ($F_1 = 0$) once $\Omega_{m}$ is determined. 

The goal now is to repeat the exercise while also fitting $F_1$ subject to the Gaussian prior (\ref{normalF}) and the new high redshift constraint $\Omega_{m} h^2/F_{\textrm{BBN}} = 0.141 \pm 0.006$.\footnote{Once again, we note that allowing $G$ to vary in CMB analysis does not greatly effect results other than increasing errors \cite{Bai:2015vca} and our high redshift prior is already generous, but of course the error may be underestimated.} However, even before performing the exercise, we have a good idea what to expect from the analytic expressions (\ref{hi}) and (\ref{w_non_min}). Any scalar profile leads to a finite value of $\alpha$, which tends to steepen the slope of $H(z)$ at $z=0$. This explains why Quintessence models with $F_1 = 0$ lower $H_0$. As the reader will observe, one can reduce $\Omega_{m}$ to compensate, but this is counteracted by any high redshift constraint on $\Omega_{m} h^2$. However, beyond $\Omega_{m}$, the scalar can now be compensated by a negative $F_1$ provided it is tolerated by the BBN prior (\ref{normalF}). In Fig. \ref{fig:nonmin} we show the output of mock fits that lower the $\chi^2$ relative to flat $\Lambda$CDM, which is defined by $(\alpha, \beta, \gamma, F_1) = (0, 0, 0, 0)$. Concretely, we repeat the same steps as outlined for Quintessence with the newly added $F_1$ parameter that is subject to a Gaussian prior (\ref{normalF}). It also contributes to the $\chi^2$ of the flat $\Lambda$CDM model when $F_1 = 0$. From 19,999 mocks, we find that roughly half, or 9,675 mocks, lead to a better fit to data. From these, we can identify 29 mocks where $H_0$ increases, while $\chi^2$ decreases. In Fig. \ref{fig:nonmin}, we separate the mocks with $\Delta \chi^2 < 0$ into those with $\Delta H_0 > 0$ (blue) and the remainder (red). One can see a strong correlation between $\Delta H_0$ and the value of $w_{\text{DE}}(z=0)$, a feature that is largely inherited from the uncoupled Quintessence model, but is in line with our analysis from section \ref{sec:warmup}. Comparing to the Quintessence mocks, we see that the probability of finding $\Delta H_0 > 0$  marginally increases to $0.15 \%$. This shows the effect of the coupling $F_1$, but it is clear that it is a small effect and throughout $\Delta H_0 < 0.14$ km/s/Mpc. As should be evident from Fig. \ref{fig:nonmin} with a coupling one can be in the phantom regime at $ z=0$, but we find that this only happens for smaller values of $\alpha$ where $|\alpha| < \sigma_{\alpha}$, so non-minimally coupled models that are pretty close to the cosmological constant $\Lambda$. For this reason, any increase in $H_0$ driven by $F_1 < 0$ is truly negligible.

\begin{figure}[htb]
  \centering
  \includegraphics[width=90mm]{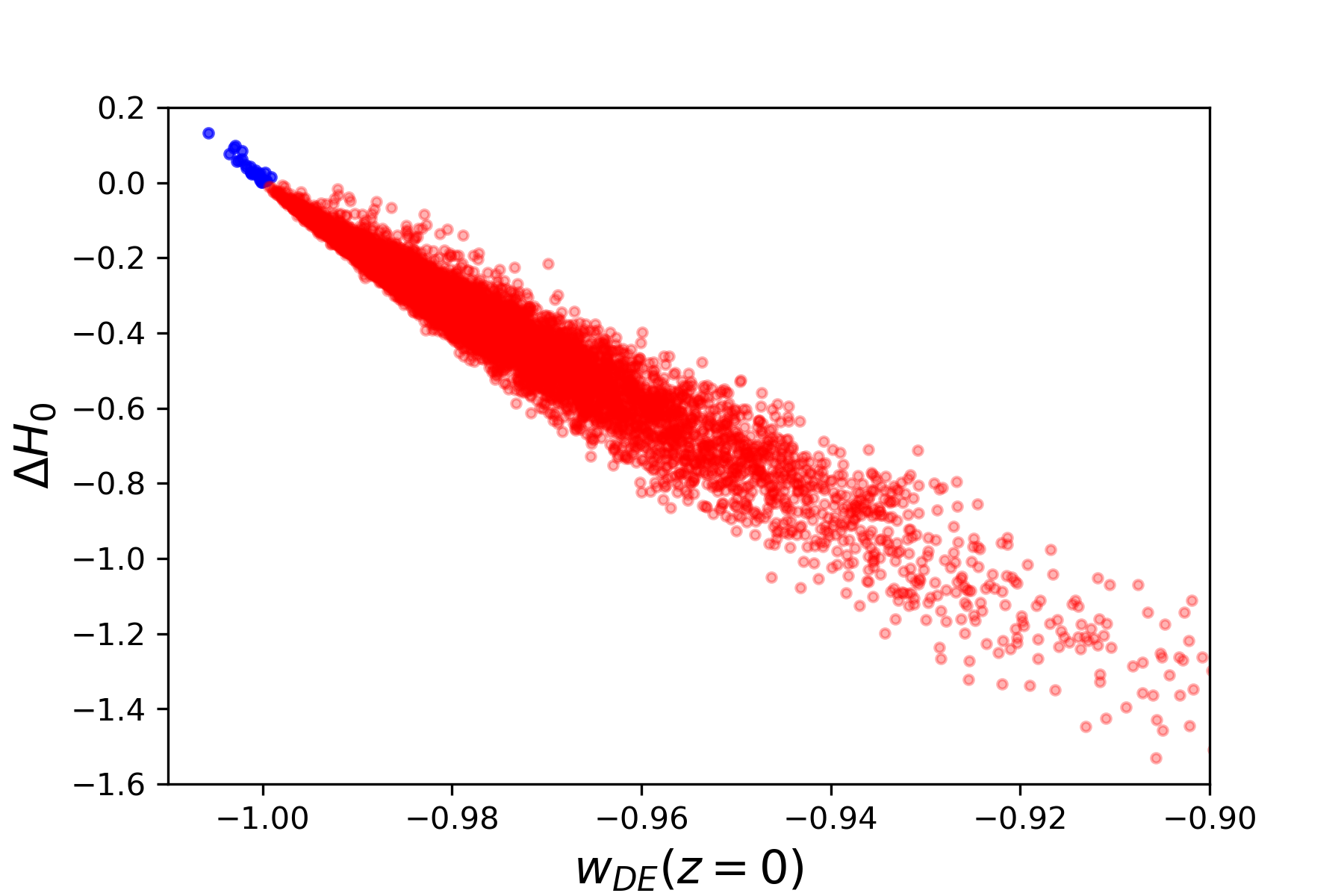}
\caption{9,675 fits to mock data and the resulting $\Delta H_0$ between the non-minimally coupled model and flat $\Lambda$CDM versus $w_{\textrm{DE}}(z=0)$ for the non-minimally coupled model. Blue separates models with $\Delta H_0 > 0$ from models with $\Delta H_0 < 0$ in red.}
\label{fig:nonmin}
\end{figure}

Our findings show that it is possible to alleviate a discrepancy with local $H_0$ values by employing a non-minimal coupling, but any effect is expected to be small. Throughout, our analysis has been conservative on a number of fronts, so the outcome may be expected. First, on the basis of a strong BBN constraint (\ref{normalF}), which is needless to say consistent with no evolution in Newton's constant, we chose linear in time evolution for the non-minimal coupling function $F(\phi)$. Obviously, one could choose another functional form for $F(\phi)$, but given there is no convincing evidence for evolution in $G$, our choice is justified. Once this simple coupling is assumed, the dynamics of the theory essentially preclude large decreases in $w_{\textrm{DE}}(z)$ as is evident from (\ref{w_non_min}), since $F_1$ makes opposing contributions at leading and subleading order in $z$. So, even at the level of the analytic expressions, one starts to see a limitation. Of course, one can alleviate $H_0$ tension marginally with respect to Quintessence, but since Quintessence is expected to generically worsen any discrepancy, this does not say much. Finally, we have not introduced a local prior on $H_0$, which means that increases in $H_0$ have to arise naturally within data that is consistent with flat $\Lambda$CDM and are not pushed by the choice of higher local $H_0$ prior. While this clearly happens for the ($w_0, w_a)$ parametrisations in section \ref{sec:warmup}, DE EFTs, even those that permit $w_{\textrm{DE}} < -1$, are evidently more rigid, and as we have seen, the scope is limited, at least within our assumptions. In that sense, our analysis agrees with \cite{Wang:2020bjk, Sakr:2021nja} that a varying Newton's constant may be a viable approach to alleviating $H_0$ tension, but this window can be expected to be severely restricted as observational constraints improve. 

It is interesting to see how our results compare with existing models in the literature. The model in  \cite{Ballesteros:2020sik}, $F(\phi) = 1 + \beta (\phi/M_{pl})^2$  alleviates $H_0$ tension, but as is clear from the quoted results therein, any improvement in $\chi^2$ versus flat $\Lambda$CDM is attributable to a $H_0$ prior. The relation of our work to  \cite{Braglia:2020iik} is less clear cut. Concretely, the authors consider the generalisation $F(\phi) = 1 + \beta (\phi/M_{\textrm{pl}})^n$ with a focus on $n =2$ and $n=4$. Naturally, as one increases $n$, any variation in $F$ is suppressed, since $\phi$ cannot be  large $|\phi| < M_{\textrm{pl}}$, so one expects less of an effect on $H_0$. While the $n=2$ model leads to a poorer fit to CMB, an observation consistent with \cite{Ballesteros:2020sik}, the $n=4$ model surprisingly improves the fit to CMB. This seems reasonable in the sense that $n=4$ represents a smaller variation in $F$ between the early and late Universe. The authors of \cite{Braglia:2020iik} once again find a reduction in $\chi^2$ when a $H_0$ prior is introduced. Notably, any reductions in $\chi^2$ for the $n=4$ model are less than the $n=2$ model, which probably implies that a local $H_0$ prior is driving the final result, since the same analysis shows that CMB prefers the $n=4$ model versus the $n=2$ model. The results of \cite{Braglia:2020iik} are also interesting for another aspect. From Table I, II and III it is clear that the radius of the sound horizon does not change much from its Planck value $r_d \approx 147$ Mpc. This tells us that the BAO scale has not changed and that increases in $H_0$ are largely down to variations in $F$. Although, $r_d$ does not feature in our analysis, since we mock up $H(z)$ and $D_{A}(z)$ data on the Planck-$\Lambda$CDM cosmology, it is implicit that our $r_d$ also adopts a Planck value.  The fact that we see such small displacements in $H_0$ in Fig. \ref{fig:nonmin} is largely down to the better quality forecasted DESI BAO data and the absence of a $H_0$ prior, since there is no conflict between variations in the Newton constant $G$ and our assumed BBN prior. The models considered in Ref. \cite{Rossi:2019lgt}  allow for $F_0 \neq 1$, so they fall outside of our assumptions, but otherwise they lead to a lower $\chi^2$. Clearly, there is still a window for non-minimally coupled Quintessence models to alleviate any discrepancy in $H_0$, but to be effective the coupling has to coherently contribute at $O(z)$ and $O(z^2)$ in the Hubble parameter so that the slope decreases. Our analysis makes it clear that from the outset, any $F(\phi)$ with close to linear evolution with time is doomed to fail to significantly increase $H_0$ and that it is unlikely there are models in this class which can do this.

\section{Kinetic Gravity Braiding} 
\label{sec:KGB}
A lot of the discussion of $H_0$ tension within the Horndeski class has focused on non-minimal couplings \cite{Rossi:2019lgt, Ballesteros:2020sik, Braglia:2020iik}. The reason being that it is easy to satisfy the stability conditions once $F(\phi) > 0$. This is no longer the case when one moves to more involved models and care is required when fitting data to ensure that one stays in a permitted range of parameter space. This is most easily done by employing MCMC and removing configurations with instabilities, see e. g. \cite{Hu:2013twa, Albuquerque:2021grl}. However, this opens up the possibility that there is a minimum for the $\chi^2$ that is not accessible to the model. Alternatively, as we do here, one can work with mock data, perform fits and simply discard any mocks that lead to violations of the stability conditions. This has the upshot that one gets an indication of how difficult the constraints are to satisfy. Nevertheless, the main point we wish to stress is that there is a large class of steadily more involved Horndeski theories, but with each generalising step, data analysis becomes steadily more complicated as the stability conditions become more involved. 

In this section we turn our attention to the Kinetic Gravity Braiding (KGB) Lagrangian \cite{Deffayet:2010qz},  
\be
\mathcal{L} = \frac{1}{2} R + K(\phi, X) + G (\phi, X) \Box \phi, 
\ee
where relative to (\ref{horndeski}) we have  redefined $G_2 (\phi, X) \rightarrow K (\phi, X)$ and $G_3 (\phi, X) \rightarrow G (\phi, X)$ to remove unsightly subscripts. In addition, we have set $M_{\textrm{pl}} =1$, so that $G_{4} = \frac{1}{2}$.\footnote{In the EFT viewpoint where we define actions up to a given (UV) scale $\Lambda$, one can be more explicit and define a dimensionless $X=-(\partial\phi)^2/\Lambda^4$.}
When $G(\phi, X) = 0$, we recover K-essence. As we have seen in section \ref{sec:warmup}, if one wants to raise $H_0$ it is imperative that $w_{\textrm{DE}}(z) < -1$ in some redshift range below the deceleration-acceleration transition redshift $z_* \approx 0.6$. For the KGB theory, the relevant equations of motion may be expressed as, 
\be\label{KGB_eqn}
\begin{split}
3 H^2 =& \rho_m + \frac{1}{2} \dot{\phi}^2+ V(\phi) - \dot{\phi}^2 ( 3 H \dot{\phi} G_{, X}- G_{, \phi}), \\
- 2 \dot{H} =& \rho_m + \dot{\phi}^2 (1 + 2 G_{, \phi} + G_{, X} ( \ddot{\phi}- 3 H \dot{\phi} )  ), 
\end{split}
\ee
where we have restricted the K-essence term to Quintessence, $K(\phi, X) = X- V(\phi)$, which is more tractable and intuitive, but we do not expect this simplification to greatly change our conclusions. We can see the importance of $G(\phi, X)$ being a function of $X$ by first setting derivatives with respect to $X$ to zero, $G_{,X} = 0$. Doing so, one observes that one can simply redefine the kinetic term, 
\be\label{KGB-to-Kessence}
\left(\frac{1}{2} + G_{,\phi}\right) \dot{\phi}^2 \rightarrow \frac{1}{2} \dot{\varphi}^2.  
\ee
In other words, when $G_{, X} = 0$, the KGB model is simply a Quintessence or K-essence model in disguise and this explains the observation in  \cite{Matsumoto:2017qil} that $w_{\textrm{DE}} < -1$ is precluded when $G_{,X} = 0$ \footnote{It is argued in \cite{Wang:2005jx}  that the interacting holographic dark energy model (HDE) permits a phantom regime, but this is countered here \cite{Kim:2005at}. Nevertheless, the minimal HDE model \cite{Li:2004rb} leads to a turning point in the Hubble parameter when confronted with observational data \cite{Colgain:2021beg}. This signals a violation of the Null Energy Condition, so minimal HDE is clearly at odds with EFT.}. Therefore, in order for the KGB Lagrangian to alleviate $H_0$ tension when confronted with observational data consistent with $\Lambda$CDM, we need to consider the general case. Once again, we will expand the Quintessence subsector perturbatively at low redshifts, which will allow us to treat it in a model agnostic fashion. 

Despite expanding $\phi$ and $V(\phi)$ following (\ref{phiV}), one sees from the equations of motion  (\ref{KGB_eqn}) that we are still confronted with a single unknown, $G(\phi, X)$. In contrast to section \ref{sec:nonmin}, the remaining function is no longer just a function of $\phi$. Furthermore, we do not have a BBN constraint that justifies reducing it to a single constant parameter.  Ideally, we would also like to expand $G(\phi, X)$, but $X$ is not guaranteed to be small. To put this comment in context, observe that $X |_{z=0} = \frac{1}{2} H_0^2 \alpha^2$, which is only less than unity if $\alpha$ is much smaller than unity, $\alpha \ll 1$. Demanding that $X < 1$ would amount to choosing $(\alpha, \beta, \gamma)$ so that one is very close to flat $\Lambda$CDM. For this reason, we will simply fix a model: 
\be
\label{G}
G (\phi, X) = g_1 X + g_2  (\phi - \phi_0) X + g_3 X^2,
\ee
where $g_i$ denote constant parameters. We have dropped any expansion solely in terms of $\phi-\phi_0$, since as explained above, this can be absorbed into a redefinition of the kinetic term. Thus, the choice (\ref{G}) represents some minimal model that allows us to quantify how higher order and terms mixing $(\phi-\phi_0)$ and $X$ affect the dynamics.  Alternatively, one could try to instead expand in $X/V(\phi)$, which is small and allows one to perform a \textit{bona fide} expansion in a small parameter, but this makes the resulting algebraic equations difficult to solve, since one starts to encounter higher order algebraic equations for $V_0$ \textit{etc}. Therefore, we simply fix $G(\phi, X)$ as in (\ref{G}). Even at this stage, one may imagine that the $g_2$ term is less relevant than the $g_1$ term, since the leading term in $\phi - \phi_0$ starts at linear order in $z$ and $z$ is small. This will indeed turn out to be true, as we will soon see. 

For any given $G(\phi, X)$ the KGB class of Horndeski theories is free of instabilities if  two conditions are met \cite{Kase:2018aps} (see also \cite{Matsumoto:2017qil}). For $K(\phi, X) = X - V(\phi)$ and $G(\phi, X)$ as in equation (\ref{G}) the stability conditions take the form \footnote{In  \cite{Kim:2004is} an attempt is made to rewrite Brans-Dicke in terms of K-essence, but one encounters a regime where the speed of sound squared is negative, $c_s^2 < 0$. Naturally, this violates stability conditions.},
\bea
\label{const1}
1 + 2 G_{, \phi} -H^2(1+z)^2 (\phi^{\prime})^2 G_{,\phi X}  - 2 H (1+z) [ H^{\prime} (1+z) \phi^{\prime} + H \phi^{\prime} + H (1+z) \phi^{\prime \prime}] G_{,X} && \nn 
+ 4 H^2 (1+z) \phi^{\prime} G_{,X} - \frac{1}{2} H^4 (1+z)^4 (\phi^{\prime})^4 G_{,X}^2 &>& 0, \\
\label{const2} 1 + 2 G_{, \phi}+H^2(1+z)^2 (\phi^{\prime})^2 G_{,\phi X} +6 H^2 (1+z) \phi^{\prime} G_{,X} + \frac{3}{2} H^4 (1+z)^4 (\phi^{\prime})^4 G_{,X}^2 &>& 0,    
\eea
When $G_{,X} = 0$, these conditions are  satisfied if $1 + 2 G_{, \phi}>0$. This is the requirement that the field redefinition \eqref{KGB-to-Kessence}, which takes us to a Quintessence model, is well-defined; Quintessence models trivially satisfy these conditions.\footnote{Flipping the sign of the Quintessence kinetic term yields ghosts, irrespective of the sign of $V(\phi)$.} This means that it is easy to fit Quintessence to data, but fitting KGB to observational data requires negotiating the conditions. In principle, this can be done by imposing the constraints in MCMC marginalisation \cite{Hu:2013twa, Albuquerque:2021grl}, but since here we generate a large number of mocks and fit each mock in turn, we will simply throw away configurations at the end that do not satisfy these conditions. As explained, this has the upshot that one gains an insight into how easy it is with mock flat $\Lambda$CDM data to evade the constraints.  

Once again, we begin by recasting (\ref{KGB_eqn}) in terms of redshift: 
\bea
3 H^2 &=& 3 H_0^2 \Omega_{m} (1+z)^3 + \frac{1}{2} H^2 (1+z)^2 (\phi^{\prime})^2  + V + 3 H^4 (1+z)^3 (\phi^{\prime})^3 G_{,X} + H^2 (1+z)^2 (\phi^{\prime})^2 G_{, \phi}, \nn
2 H  H^{\prime} &=& 3 H_0^2 \Omega_{m} (1+z)^2 + 
H^2(1+z) (\phi^{\prime})^2 \biggl( 1 + 2 G_{, \phi} + H(1+z) \left[ H^{\prime} (1+z) \phi^{\prime} + 4 H \phi^{\prime} + H (1+z) \phi^{\prime \prime} \right] G_{, X} \biggr), \nonumber
\eea
where  $\rho_{m} = 3 H_0^2 \Omega_{m} (1+z)^3$. One now proceeds to solve order by order. At leading order, one can solve for the constant component of the potential $V_0$ and the linear in $z$ term in the Hubble parameter: 
\bea
\label{V0h1}
V_0 H_0^{-2} &=& 3 (1- \Omega_{m}) - \frac{1}{2} \alpha^2 (1+ g_2 H_0^2 \alpha^2) - 6 (1- \Delta), \nn
h_1 &=& \frac{1}{2} \alpha^2 (1+ g_2 H_0^2 \alpha^2) + \frac{3}{2} \Omega_{m} + \frac{(1- \Delta)}{2 \Delta} \left( \alpha^2 (1+ g_2 H_0^2 \alpha^2)+ 4 \frac{\beta}{\alpha} + 8 + 3 \Omega_{m} \right), 
\eea
where  $\Delta := 1 - \frac{1}{2} g_1 H_0^2 \alpha^3 - \frac{1}{2} g_3 H_0^4 \alpha^5$. The first observation is that setting $g_i = 0$ we recover the expressions in Ref. \cite{Banerjee:2020xcn}. Next, we see that despite the scalar increasing the slope of $H(z)$ for fixed $\Omega_{m}$, this can be counteracted by $\Delta > 1$ ($g_1 \alpha < 0$) and $g_2 < 0$. Interestingly, the linear and quadratic $X$ terms in (\ref{G}) enter $h_1$ through $\Delta$, which means that they play more or less the same role, at least at linear order, and as we have seen in section \ref{sec:warmup}, it is the lowest order terms that are most relevant. Moreover, it is easy to check that for any higher order power of $X^n, n \in \mathbb{N}$ in (\ref{G}) that they can all be absorbed into a single $\Delta$, thereby underscoring the redundancy at leading order. Given the redundancy between $g_1$ and $g_3$, it is enough to set one of them to zero, so we set $g_3 = 0$ and focus exclusively on $g_1 $ and $g_2$.  

Unfortunately, beyond this order expressions quickly become unwieldy even with $g_3 = 0$, so further simplifications are in order. Thus, we  terminate our expansion at second order in the Hubble parameter in contrast to third order in  Ref. \cite{Banerjee:2020xcn} and section \ref{sec:nonmin}. With a focus on the flat $\Lambda$CDM model, we illustrate how much of an approximation this entails in appendix \ref{sec:approx}. At second order, we find the expressions, 
\bea
V_1 H_0^{-2} &=&  - \frac{1}{2} \alpha^3 - 2 \beta + 2 \alpha - \frac{3}{2} \alpha \Omega_m - \frac{1}{2} g_2 {\alpha^2}  H_0^2 \left(3 \alpha^3+\alpha  (6 \Omega_m {+4})+{8} \beta +2 g_2 {\alpha^5}  H_0^2 \right) \nn
&-& \frac{(1-\Delta)}{2 \alpha^2 \Delta} {\biggl[ \alpha^3 (3 \Omega_m +26)+4 \alpha^2 \beta -6 \alpha  (26 \Delta -9 \Omega_m -24)-24 \beta  (\Delta -3)+2 \alpha^9 g_2^2 H_0^4+3 \alpha^7 g_2 H_0^2 } \nn 
 &+&{ \alpha^5 [ 1+2 g_2 H_0^2 (3 \Omega_m +17) ] +8 \alpha^4 \beta  g_2 H_0^2 \biggr], } \nonumber
\eea
\bea
h_2 &=& \frac{1}{8} \alpha^4 + \frac{1}{4} \alpha^2 + \alpha \beta + \frac{3}{8} \Omega_m (4 - 3 \Omega_m) 
+ \frac{g_2 H_0^2 {\alpha^3} }{8} {[5 \alpha^3 + 4 g_2 H_0^2 \alpha^5 + 20 \beta + \alpha (14 + 9 \Omega_{m})]} \nn 
&+& \frac{(1-\Delta)}{8 \alpha^2 \Delta^3} {\biggl[ 2 \alpha^4 \left(\Delta^2+3 \Delta +6 \Omega_m +16\right)+4 \alpha^3 \beta  \left(2 \Delta^2+3 \Delta+4\right)  +32 \beta ^2 \left(-\Delta ^2+\Delta +1\right) }\nn
&+& {\alpha^2 \left[ \Delta^2 \left(-9 \Omega_m^2+12 \Omega_m -80\right)-\Delta  \left(9 \Omega_m^2+30
   \Omega_m +16\right)+2 (3 \Omega_m +8)^2\right] } \nn
&+& {4 \alpha  \left[ \beta  \left(-16 \Delta^2+3 \Delta  \Omega_m +14
   \Delta +12 \Omega_m +32\right)+6 \gamma  \Delta ^2\right]+2
   \alpha ^{10} \left(2 \Delta ^2+2 \Delta +1\right) g_2^2 H_0^4 }\nn
&+& {\alpha ^8 \left(5 \Delta ^2+5\Delta +4\right) g_2 H_0^2 +4 \alpha^5 \beta  \left(5 \Delta^2+6 \Delta +4\right) g_2 H_0^2}\nn
&+& {\alpha ^6 \left(\Delta^2+\Delta +2 +g_2 H_0^2 \left(\Delta
   ^2 (9 \Omega_m +14)+\Delta  (9 \Omega_m +30)+4 (3 \Omega_m +8)\right)\right) \biggr]. }\nonumber
\eea
Clearly, when $g_1 = g_2 = 0$ we recover earlier expressions \cite{Banerjee:2020xcn}. Now, we could proceed to determine $V_3$ and $h_3$, but the expressions are already pretty intimidating, at least relative to the non-minimal coupling case, so as explained, we simply terminate at second order in $z$. Finally, just as in  \cite{Banerjee:2020xcn}, it is instructive to record the DE EoS, 
\bea
w_{\textrm{DE}} &=&  -1 + \frac{\dot{\phi}^2 \left(1+ 2 G_{, \phi} + [\ddot{\phi} - 3 H \dot{\phi}] G_{,X} \right)}{\frac{\dot{\phi}^2}{2} + V - \dot{\phi}^2 ( 3 H \dot{\phi} G_{,X}-G_{,\phi})}, \\
&=&  -1 + \frac{1}{3 (1 - \Omega_{m})} \left[ \alpha^2 (1+ g_2 H_0^2 \alpha^2) + \frac{(1-\Delta)}{\Delta} \left( \alpha^2 (1+ g_2 H_0^2 \alpha^2 ) + 4 \frac{\beta}{\alpha} + 8 + 3 \Omega_{m} \right)  \right] +  O(z). \nonumber
\eea
Unsurprisingly, this expression also becomes intractable and not very insightful beyond leading order. That being said, even at leading order, there is a nugget of information to be gained. The first observation is that a profile for the scalar, i. e. $\alpha$, will always lead to an increase in $w_{DE}(z)$ at $z = 0$, however as noted above with $h_1$, mock $\Lambda$CDM data can attempt to counter this through $g_2 < 0$ and $\Delta > 1$. That being said, at this stage it is not immediately clear which mechanism the data will exploit to do this, so we defer any discussion until after we have performed some mock fits. 

\begin{figure}[htb]
\centering
\begin{tabular}{cc}
\includegraphics[width=75mm]{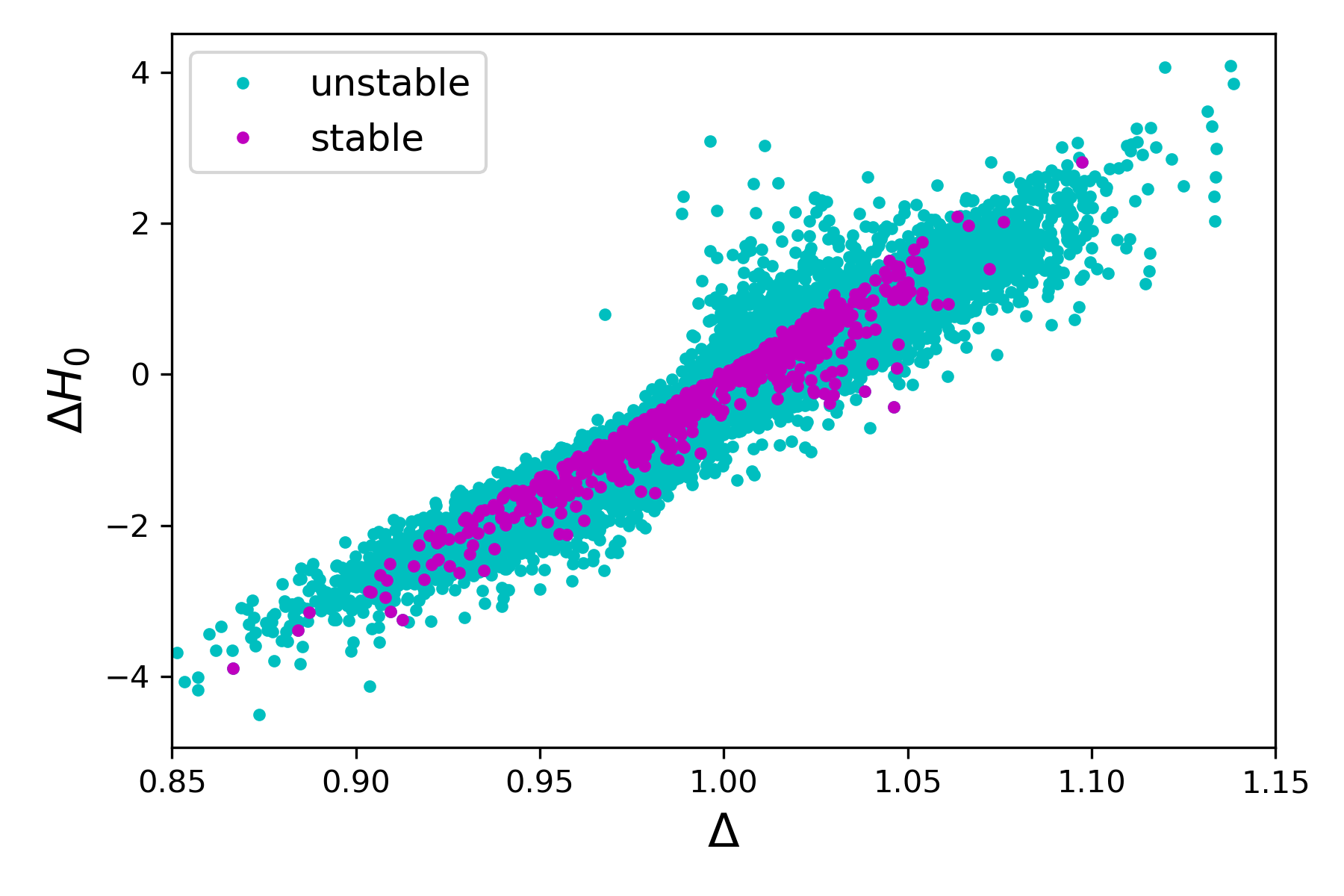}&
\includegraphics[width=75mm]{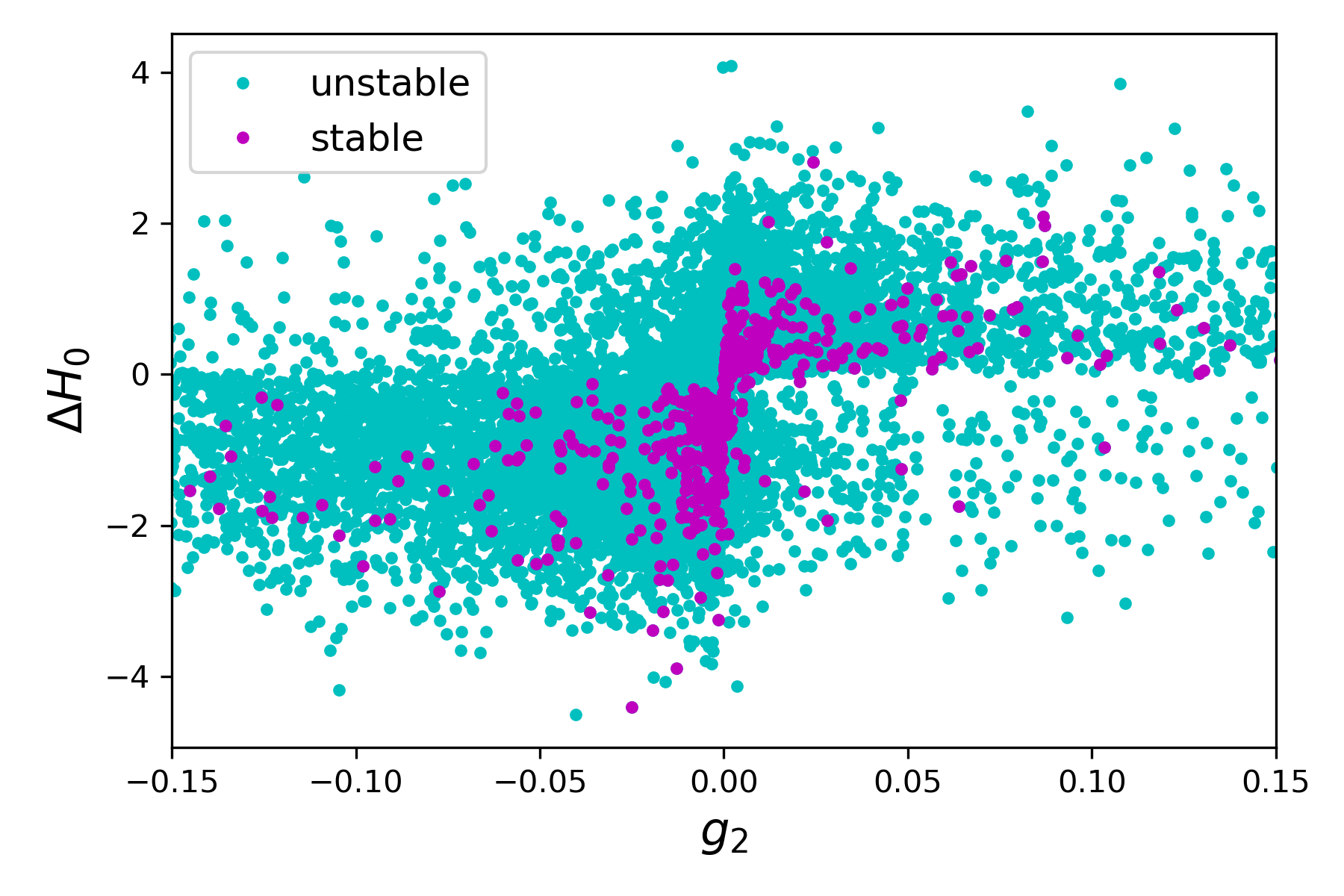}
\end{tabular}
\caption{These plots depict how much one can increase $H_0$ as we move in the parameter space of perturbative KGB model.}
\label{fig:KGB1} 
\end{figure}

Let us now turn our attention to the mocks, which we perform in an analogous fashion to section \ref{sec:warmup}. In particular, we employ the same high redshift constraint on the combination $\Omega_m h^2$, we generate random scalar profiles with standard deviations $(\sigma_{\alpha}, \sigma_{\beta}, \sigma_{\gamma}) = (0.18, 0.12, 0.06)$ subject to (\ref{constraints}), while fixing flat priors for $g_i$, namely $ -0.05 < g_1 < 0.05$ and $-0.2 < g_2 < 0.2$. Over a large number of mocks, we find that a number of best fits saturate our bounds, but on the whole the bounds do not impact the results. We run and fit 20,214 mocks, from which we identify 19,443 mocks where the two additional parameters $g_1$ and $g_2$ succeed in compensating the scalar profile defined by ($\alpha, \beta, \gamma)$ enough to improve the fit to the data by lowering the $\chi^2$. However, once we impose the constraints ensuring stability (\ref{const1}) and (\ref{const2}), this reduces the number of valid mocks and corresponding best fits to 646. This represents $3.2 \%$ of the original mocks, which increases to $3.3\%$ if we neglect mocks where the KGB model fails to fit the data well. This serves to highlight the restrictions that the stability conditions place on parameter space. It should be noted that we are only imposing the constraints (\ref{const1}) and (\ref{const2}) where data exists in the range $0.05 \leq z \leq 0.95$ in line with Table \ref{DESI_table}. In Fig. \ref{fig:KGB1}, we show that increases in $H_0$ relative to flat $\Lambda$CDM are primarily driven by $\Delta$. Moreover, we find that $g_2$ plays a limited role, and contrary to our initial expectations, increases in $H_0$, i. e. $\Delta H_0 > 0$ are correlated with increases in $g_2$, and \textit{vice versa}. This shows the clear preference for the $g_1$ term in (\ref{G}), since analytically $g_2 > 0$ cannot reduce the slope of the Hubble parameter, thereby increasing $H_0$. 

\begin{figure}[htb]
\centering
\includegraphics[width=90mm]{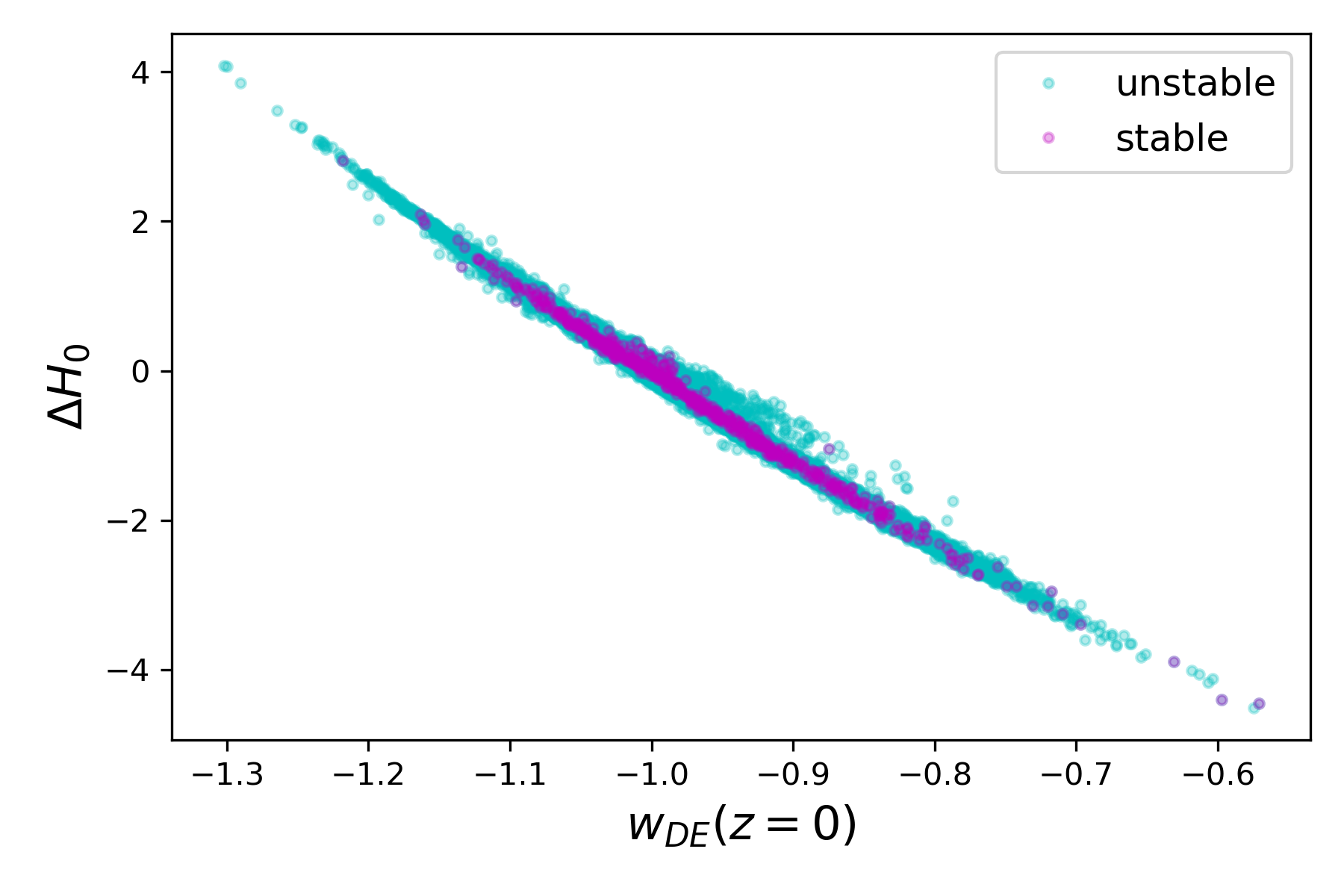}
\caption{Increases in $H_0$ versus $w_{\textrm{DE}}(z=0)$ for the KGB model. The stable configurations in magenta explore a restricted parameter space leading to less pronounced increases in $H_0$.}
\label{fig:KGB2} 
\end{figure}

Finally, given our results in section \ref{sec:warmup}, it is interesting to see how increases in $H_0$ are correlated with $w_{\textrm{DE}}(z=0) < -1$. In Fig. \ref{fig:KGB2} we show that tangible increases in $H_0$ involve sizable excursions into the phantom regime at $z=0$. Moreover, as is clear from the plot, the stability conditions (\ref{const1}) and (\ref{const2}), which ensure absence of ghosts and speed of sound $c_s^2>0$, do not allow $1+w_{\textrm{DE}}$ and $\Delta Y(z_*)$ to become too negative. It is worth noting that the earlier \textit{phenomenological} ($w_0, w_a$) parametrisations from section \ref{sec:warmup} can access these points in parameter space and this serves to underscore the limitation of typical DE EFTs, when increases in $H_0$ are required. In essence, DE EFTs are not only analytic, as $(w_0, w_a)$ parametrisations are, but they are also restricted by stability conditions. It is these additional conditions that reduces the parameter space.  


\section{Discussion} 
We introduced our study by emphasising that $\Lambda$ is a phenomenological parameter that simply quantifies our ignorance of the DE sector. Since EFT is the language of physics, ultimately one would expect $\Lambda$ to be absorbed in an EFT description. However, this expectation seems to run contrary to the observation that local $H_0$ determinations are universally biased larger than Planck-$\Lambda$CDM. In particular, as has been demonstrated \cite{Vagnozzi:2018jhn, Vagnozzi:2019ezj, Alestas:2020mvb, Banerjee:2020xcn}, DE models in the Quintessence regime, $w_{\textrm{DE}}(z)>-1$, notably the traditional EFT regime, are expected to lower $H_0$, thereby exacerbating any $H_0$ discrepancy/tension. Naively, there seems to be some apparent conflict between observed higher local $H_0$ determinations and DE EFT, at least in the traditional XCDM paradigm, where X is a generic late DE model described by a healthy EFT. Clearly this has far-reaching consequences, since some experimental proposals for dark energy implicitly rest on the assumptions of a space pervading scalar field \cite{Vagnozzi:2021quy, Zhang:2021ygh}.

Section \ref{sec:warmup} picks up this thread within simple parametrisations for an evolving DE EoS, $w_{\textrm{DE}}(z) = w_0 + w_a f(z)$, where $f(z)$ is a function satisfying $f(z=0) = 0$, while still being finite at high redshifts. Within this setting, once one has evolution, there is little to preclude an EoS that crosses the phantom divide $w_{\textrm{DE}}(z)=-1$, and an obvious implication of the observation that $w_{\textrm{DE}}(z) > -1$ lowers $H_0$ relative to the cosmological constant $\Lambda$ \cite{Vagnozzi:2018jhn, Vagnozzi:2019ezj, Alestas:2020mvb, Banerjee:2020xcn} is that models with $w_{\textrm{DE}}(z) < -1$ can raise $H_0$. Indeed, recent papers \cite{Heisenberg:2022lob, Heisenberg:2022gqk} give an analytic proof that $w_{\textrm{DE}}(z) < -1$ is a prerequisite for alleviating $H_0$ tension. However, this still leaves open the question what is the most relevant redshift $z$ where $w_{\textrm{DE}}(z) < -1$? Our analysis in section \ref{sec:warmup} provides an answer within ($w_0, w_a)$ parametrisations of the EoS. Concretely, we introduced an integrated (normalised) DE density, $Y(z)$, defined through to the deceleration-acceleration transition $z_*$, and showed over a large number of flat $\Lambda$CDM mocks that {$H_0$ is guaranteed to increase, thereby alleviating tension, if \textit{both} $\Delta w_0 = w_0 + 1 < 0$ \textit{and} $Y(z_*) < z_*$}. These conditions imply $w_{\textrm{DE}}(z) < -1$ must hold somewhere between $z=0$ and the transition redshift $z_* \approx 0.6$. {In contrast, DE models satisfying $\Delta w_0 > 0$ and $Y(z_*) > z_*$ lower $H_0$. Finally, one has models where $\Delta w_0 \cdot \Delta Y(z_*) < 0$, which either lead to small increases or decreases within $1 \sigma$, whereas the most pronounced displacements in $H_0$ are reserved for models with $\Delta w_0 \cdot \Delta Y(z_*) > 0$. In addition}, our analysis based on flat $\Lambda$CDM mocks clearly demonstrates that $w_0$ is the most relevant parameter (see Fig. \ref{fig:w0wa}). Naturally, $w_0 < -1$ guarantees $w_{\textrm{DE}}(z) < -1$ in some vicinity of $z=0$, but it only implies $Y(z_*) < z_*$ when $w_a = 0$ (no evolution). 
It is also worth noting that $w_0$ and $w_a$ are anti-correlated (Fig. \ref{fig:w0wa}), since the mock data is consistent with flat $\Lambda$CDM by construction.   

Throughout we have employed BAO mocks based on DESI forecasts \cite{DESI:2016fyo}. Obviously, one is free to combine with other data sets, but our conclusions are not expected to change. The rationale for employing BAO is that BAO leads to strong constraints on $H(z)$ and $D_{A}(z)$ at effective redshifts. In short, not only is one averaging or coarse graining over redshifts, but one is extracting quantities that are largely only sensitive to the combination $\Omega_{m} H_0^2$, e. g. $H(z), D_{A}(z)$, in line with equation (\ref{hubble}). For this reason, any evolution of $H_0$ with redshift \cite{Wong:2019kwg, Millon:2019slk, Krishnan:2020obg, Dainotti:2021pqg, Dainotti:2022bzg},\footnote{This evolution, if real, could be driven by increases in $\Omega_{m}$ within the flat $\Lambda$CDM model. Concretely, when one fits Pantheon SN \cite{Pan-STARRS1:2017jku} with a low redshift cutoff $z_{\textrm{min}}$ so that SN below $z_{\textrm{min}}$ are removed, one finds that the central value of $H_0$ decreases in line with Ref. \cite{Dainotti:2021pqg}, while $\Omega_{m}$ increases, with increasing $z_{\textrm{min}}$. The statistical significance of the feature is approximately $1 \sigma$.} which it should be stressed is a necessary condition for $H_0$ tension to have a resolution within FLRW \cite{Krishnan:2020vaf, Krishnan:2022fzz} (see also \cite{Mawas:2021tdx}), is expected to be washed out. In this sense, BAO represents a conservative data set that is more likely to recover results consistent with Planck-$\Lambda$CDM. Nevertheless, discrepancies have been reported in the literature \cite{BOSS:2014hwf, duMasdesBourboux:2020pck,  DES:2021esc} and this motivated us to check the results in section \ref{sec:warmup} were robust by limiting redshift ranges. Moreover, as demonstrated in  \cite{Colgain:2021pmf}, $(w_0, w_a)$ EoS parametrisations suffer from a degree of arbitrariness, so it is imperative to make sure that statements hold across different models (see \cite{Yang:2021flj}), which we have done.  

As explained in the text, our analysis in section \ref{sec:warmup} provides another perspective on the fact that Quintessence models \cite{Ratra:1987rm,Wetterich:1987fm} and K-essence models \cite{Armendariz-Picon:1999hyi, Chiba:1999ka, Armendariz-Picon:2000ulo} predict \textit{lower} values of $H_0$ relative to $\Lambda$. This places them at odds with local $H_0$ determinations, which should be enough to rule out these models, and indeed any DE EFT with $w_{\textrm{DE}}(z) > -1$ (see also \cite{Heisenberg:2022lob, Heisenberg:2022gqk}), as viable late-time DE EFTs. Going further, the fact that the largest increases in $H_0$  are driven by models with $w_{\textrm{DE}}(z=0) < -1$ suggests that coupled Quintessence \cite{Amendola:1999er}, where $w_{\textrm{DE}}(z=0) > -1$ \cite{Das:2005yj}, will be less effective in inducing the significant increases in $H_0$ required if local $H_0$ determinations converge to the higher values permitted within the FLRW paradigm, namely $H_0 \lesssim 73$ km/s/Mpc at $2 \sigma$ \cite{Krishnan:2021dyb, Vagnozzi:2021tjv}. The restriction is already evident from recent analysis \cite{Gomez-Valent:2020mqn}. Of course, all of these models can be resurrected as EDE models \cite{Poulin:2018cxd, Niedermann:2019olb}, but this does not alter the conclusion that, even in such a scenario, $\Lambda$ is expected to maximise $H_0$ over the simplest EFTs in the Horndeski class. {In summary, the survival of DE EFT now hangs by an early universe thread, which already constitutes a paradigm shift in DE EFT. \footnote{See Ref. \cite{Scherrer:2022umm} for a recent Quintessence study where new early universe physics is invoked as physical motivation.}}

However, Quintessence and K-essence are simply the tip of the Horndeski iceberg, and since one can find EFTs that cross the phantom divide \cite{DeFelice:2011bh, Matsumoto:2017qil}, the latter part of our paper conducted a preliminary, yet reasonably general investigation by employing a model agnostic approach for the Quintessence subsector. One relevant question is whether a non-minimal coupling to gravity can help alleviating $H_0$ tension? Concretely, section \ref{sec:nonmin} addresses this question within the assumption that the non-minimal coupling $F(\phi)$ varies linearly with cosmic time subject to a recent BBN constraint \cite{Alvey:2019ctk}. As we have seen, the slope $F_1$ contributes to the Hubble parameter in a conflicted manner; if $F_1$ flattens the slope of the Hubble parameter at leading order in redshift $z$, it increases the slope at subleading order. For this reason, one would expect increases in $H_0$ due to linear evolution of $F(\phi)$ with cosmic time to be marginal, and this is indeed what we found. Even though the outlook may not be good \cite{Wang:2020bjk,Sakr:2021nja}, at a technical level it is imperative to identify models that increase $H_0$ without relying heavily on a local $H_0$ prior, e. g. \cite{Ballesteros:2020sik, Braglia:2020iik}. Finally, we note that increasing $H_0$ requires $\dot{F} > 0$ $(F_1 < 0$), which implies the Newton's constant must decrease in the late universe in line with the findings of Ref. \cite{Heisenberg:2022lob, Heisenberg:2022gqk}. This can also be seen from the equations of motion (\ref{EOM}) and (\ref{hubble}), once one appreciates that increases in $F(\phi)$ reduce the DE density $X(z)$ and thus lowering $H_0$ in line with section \ref{sec:warmup}. 

In the final section \ref{sec:KGB}, we looked at Kinetic Gravity Braiding models \cite{Deffayet:2010qz}. As late-time DE models, these are expected to fall under the analysis of Ref. \cite{Krishnan:2021dyb, Cai:2021weh}, which shows that cosmological data including BAO restricts late universe modifications within Einstein gravity to central values below $H_0 = 70$ km/s/Mpc. Naturally, this still places them in tension with the recent SH0ES result \cite{Riess:2021jrx}. Concretely, we found that $X$-dependent contributions to the braiding function $G(\phi, X)$ alleviate $H_0$ tension in a meaningful way compared to models with non-minimal coupling. This difference can be traced analytically to the fact that any $X^n$, $n > 0 \in \mathbb{N}$, contribution to $G(\phi, X)$ can coherently flatten $H(z)$ at both leading and subleading order in $z$ for a sizable class of Quintessence models. As we have shown, this leads to tangible increases in $H_0$ that correlate well with a phantom EoS at $z=0$, $w_{\textrm{DE}}(z=0) < -1$. {We have employed the same mocking procedure throughout, so one is free to compare Fig. \ref{fig:KGB2} with Fig. \ref{fig:lowz_Taylor}, since we have used data in the same redshift range. Doing so, one is comparing the output of $\sim 20,000$ mocks, including unstable configurations, in Fig. \ref{fig:KGB2} with $\sim 10,000$ mocks in Fig. \ref{fig:lowz_Taylor}. One notes that $\Delta H_0$ is less for the EFT and this gets worse as one restricts attention to stable mocks, admittedly with greatly decreased numbers, so comparison is less meaningful. Ultimately, EFT is more structured than ($w_0, w_a)$ parametrisations, since one is not only constrained by analyticity, but also stability conditions, so smaller displacements in $H_0$ are expected. As a result, if one can resolve $H_0$ tension within a given ($ w_0, w_a)$ parametrisation, only then does it make sense to study the problem within EFT.}

Finally, while still less significant than the $H_0$ tension, there are many papers trying to address $S_8$ and the $H_0$ tensions, see e.g. \cite{Schoneberg:2021qvd} and references therein. The lore is that late DE models which alleviate $H_0$ typically exacerbate $S_8$. The recent papers \cite{Heisenberg:2022lob, Heisenberg:2022gqk} explicitly verify this lore in a general setting. Based on these results, it is reasonable to expect that the $S_8$ considerations should only strengthen our results here that the DE EFT framework is less likely to hold the answer to the cosmic tensions, but one is always free to venture theoretically beyond EFT, e. g. \cite{Giusti:2021shf,vanPutten:2021hlu,Cai:2021wgv}. In the big picture, if attempts to alter the BAO scale through EDE or equivalent are discredited \cite{Hill:2020osr,Ivanov:2020ril,DAmico:2020ods, Niedermann:2020qbw, Murgia:2020ryi, Smith:2020rxx, Jedamzik:2020zmd, Lin:2021sfs, Vagnozzi:2021gjh, Herold:2021ksg}, one is confronted with a  $\Lambda$ that does not appear to admit an EFT description. If confirmed, this in itself is an extremely profound insight into the cosmological constant $\Lambda$.

\section*{Acknowledgements}
We thank Aritra Banerjee, Gansukh Tumurtushaa and Lu Yin for discussions on related topics. In addition, we thank Roberto Casadio, Hongsu Kim, Maurice van Putten, Shao-Jiang Wang, Wen Yin and Zhen Zhang for correspondence.  E\'OC was supported by the National Research Foundation of Korea grant funded by the Korea government (MSIT) (NRF-2020R1A2C1102899). MMShJ would like to acknowledge SarAmadan grant No. ISEF/M/400121. BHL,WL,ST were supported by the Basic Science Research Program (2020R1A6A1A03047877) of the National Research Foundation of Korea funded by the Ministry of Education through Center for Quantum Spacetime (CQUeST) of Sogang University.\\
BHL(2020R1F1A1075472) and WL(NRF-2016R1D1A1B01010234) were supported by Basic Science Research Program through the National Research Foundation of Korea funded by the Ministry of Education.
\newpage

\appendix 

\section{DESI BAO forecasts}
\label{sec:DESI}
We reproduce the forecasted DESI errors for $H(z)$ and $D_{A}(z)$ in Table \ref{DESI_table}. 

\begin{table}[htb]
\centering
\begin{tabular}{c|c|c}
$z$ & $\sigma_{H}/H$ (\%) & $\sigma_{D_{A}}/D_{A}$ (\%) \\
\hline
$0.05$ & $12.10$ & $6.12$ \\
$0.15$ & $4.66$ & $2.35$ \\
$0.25$ & $2.97$ & $1.51$ \\
$0.35$ & $2.44$ & $1.32$ \\
$0.45$ & $3.69$ & $2.39$ \\ 
$0.65$ & $1.50$ & $0.82$ \\ 
$0.75$ & $1.27$ & $0.69$ \\ 
$0.85$ & $1.22$ & $0.69$ \\ 
$0.95$ & $1.22$ & $0.73$ \\ 
\hline
$1.05$ & $1.37$ & $0.89$ \\ 
$1.15$ & $1.39$ & $0.94$ \\ 
$1.25$ & $1.39$ & $0.96$ \\ 
$1.35$ & $2.02$ & $1.50$ \\ 
$1.45$ & $2.13$ & $1.59$ \\ 
$1.55$ & $2.52$ & $1.90$ \\ 
$1.65$ & $3.80$ & $2.88$ \\ 
$1.75$ & $6.30$ & $4.64$ \\ 
$1.85$ & $6.39$ & $4.71$ \\ 
$1.96$ & $3.42$ & $3.35$ \\ 
$2.12$ & $2.48$ & $2.43$ \\ 
$2.28$ & $2.63$ & $2.72$ \\ 
$2.43$ & $2.82$ & $3.07$ \\ 
$2.59$ & $3.08$ & $3.57$ \\ 
$2.75$ & $3.44$ & $4.24$ \\ 
$2.91$ & $3.96$ & $5.26$ \\ 
$3.07$ & $4.62$ & $6.60$ \\ 
$3.23$ & $5.70$ & $8.86$ \\ 
$3.39$ & $7.72$ & $13.05$ \\ 
$3.55$ & $11.09$ & $19.85$ \\ 
\end{tabular} 
\caption{Forecasted DESI percentage errors for $H(z)$ and $D_{A}(z)$ assuming 14,000 deg$^2$ sky coverage \cite{DESI:2016fyo}. We will primarily be interested in the entries below $z =1$.}
\label{DESI_table}
\end{table} 
\section{Analytic Approximations} 
\label{sec:approx}
Here we demonstrate that the approximations used in sections \ref{sec:KGB} and \ref{sec:nonmin} are under control. To do so, we define the fractional error in $H(z)$ and $D_{A}(z)$ as 
\be
\Delta H(z) = 1 - H(z)_{\textrm{approx}}/H(z)_{\textrm{exact}}, \quad \Delta D_{A} (z) = 1 - D_{A}(z)_{\textrm{approx}}/D_{A}(z)_{\textrm{exact}}, 
\ee
where the exact expressions for $H(z)$ and $D_{A}(z)$ read 
\be
H(z)_{\textrm{exact}} = H_0 \sqrt{ 1- \Omega_{m} + \Omega_{m} (1+z)^3}, \quad D_{A}(z)_{\textrm{exact}} = \frac{c}{(1+z)} \int_{0}^{z} \frac{1}{H(z^{\prime})_{\textrm{exact}}} \dd z^{\prime}
\ee
Expanding $H(z)_{\textrm{exact}}$ to third order, one has 
\be
H(z)_{\textrm{approx}} = H_0 \left( 1 + \frac{3}{2}  \Omega_{m} z + \frac{3}{8} \Omega_{m} (4 - 3 \Omega_{m}) z^2 + \frac{1}{16} \Omega_{m} (8 - 36 \Omega_{m}+27 \Omega_m^2) z^3 + O (z^4) \right). 
\ee
One then defines $D_{A}(z)_{\textrm{approx}}$ in an analogous fashion to above, but one should perform the integral numerically without first expanding $1/H(z)_{\textrm{approx}}$ in $z$. Note, it is common for one to expand $D_{A}(z)_{\textrm{approx}}$ in the literature, e. g. \cite{Visser:2003vq}, but in inverting $H(z)_{\textrm{approx}}$ one makes the approximation unnecessarily worse. This can be avoided by simply numerically integrating. We illustrate the approximations for expansions that terminate at second and third order in Fig. \ref{2ndapprox} and Fig. \ref{3rdapprox}, respectively. One can see that any error in $D_{A}(z)$ is negligible. This is expected since even if the error in $H(z)$ grows with increasing $z$, these terms make a smaller contribution to the integral. The error in $H(z)$ is more serious, but is never more than $2 \%$ and $3 \%$ for values of $\Omega_{m}$ close to the Planck value. 

\begin{figure}[htb]
   \centering
\begin{tabular}{cc}
\includegraphics[width=75mm]{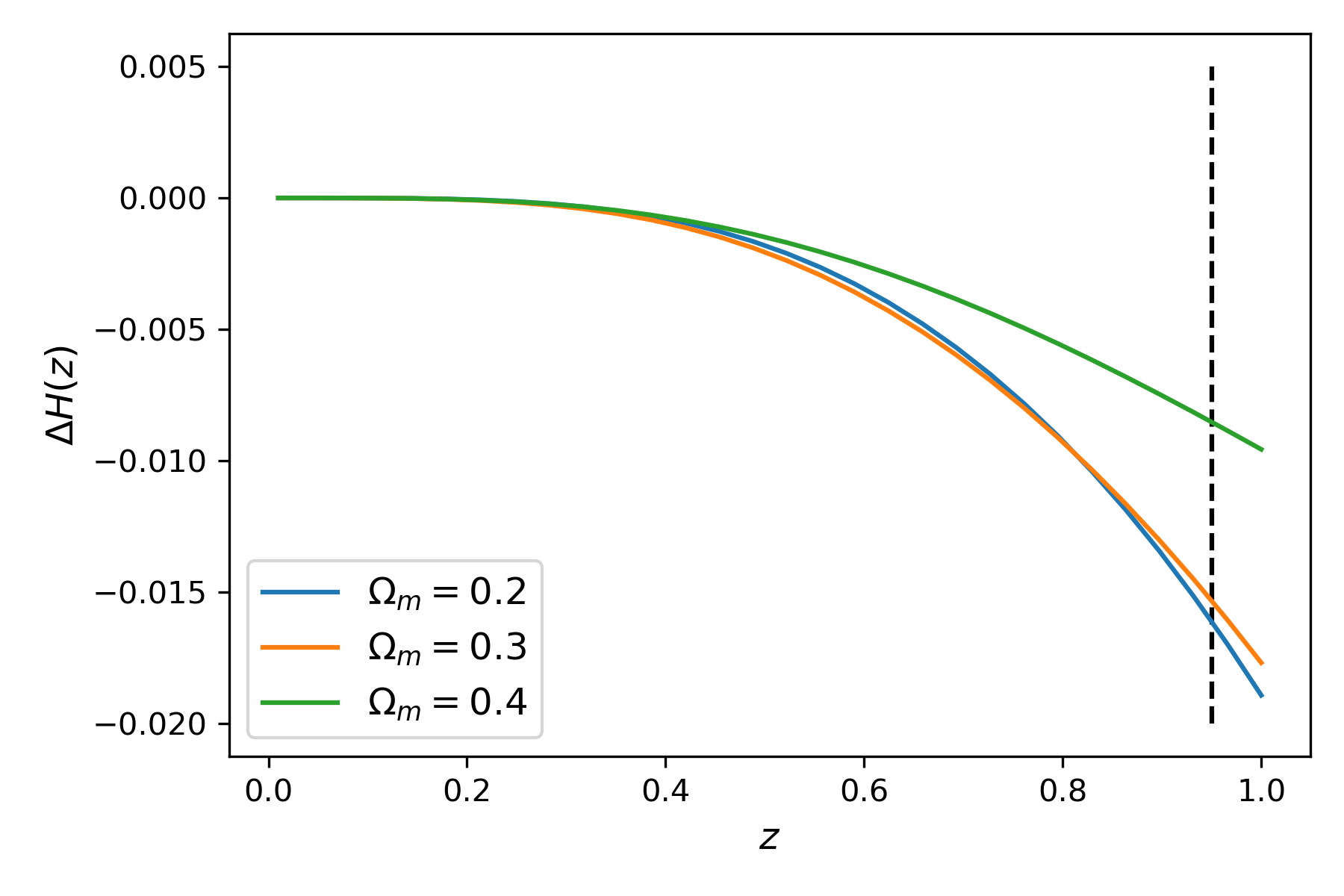}&
\includegraphics[width=75mm]{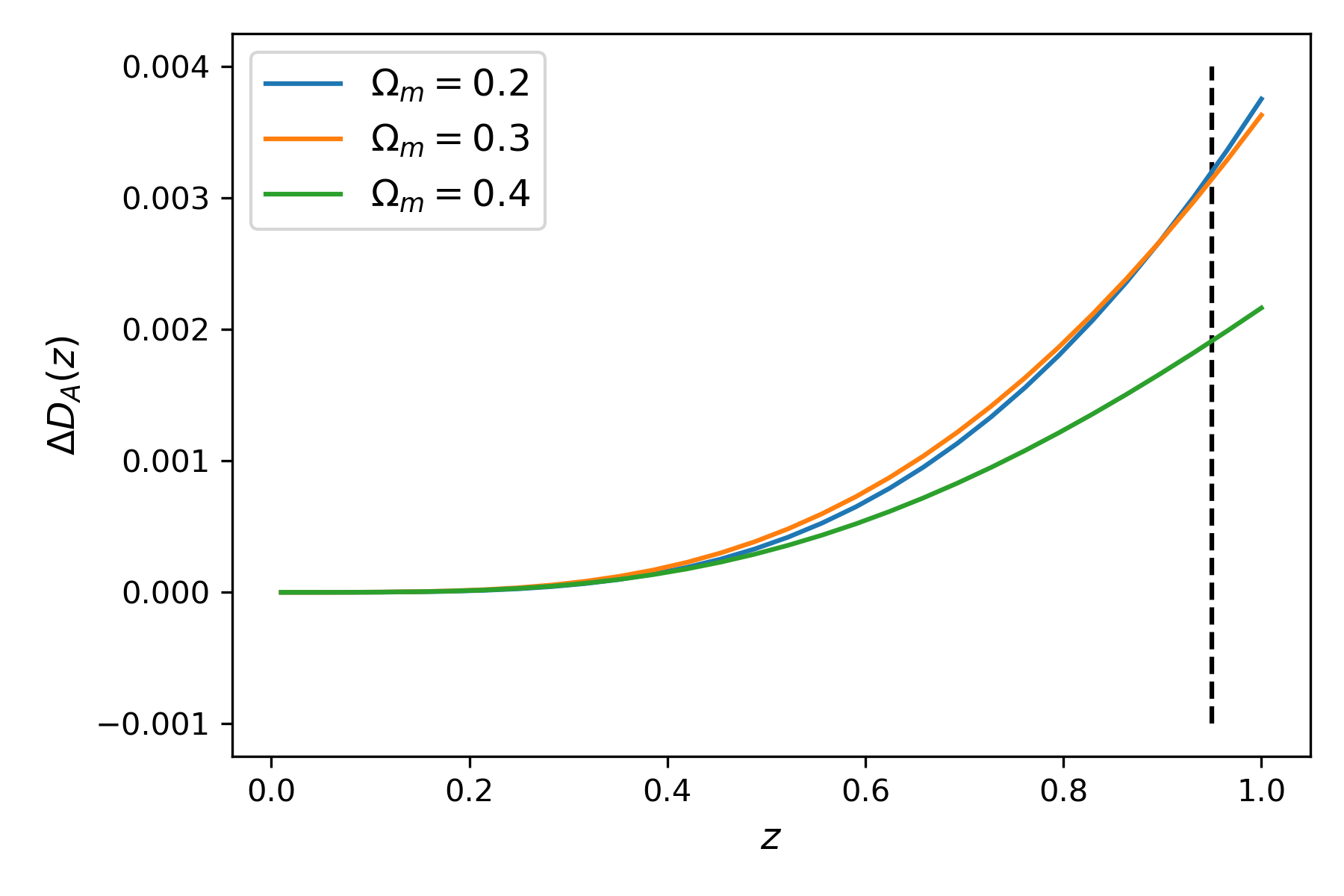}
\end{tabular}
\caption{Fractional error in $H(z)$ and $D_{A}(z)$ approximations at third order. Dashed lines denote the highest redshift in our sample.}
\label{2ndapprox} 
\end{figure}

\begin{figure}[htb]
   \centering
\begin{tabular}{cc}
\includegraphics[width=75mm]{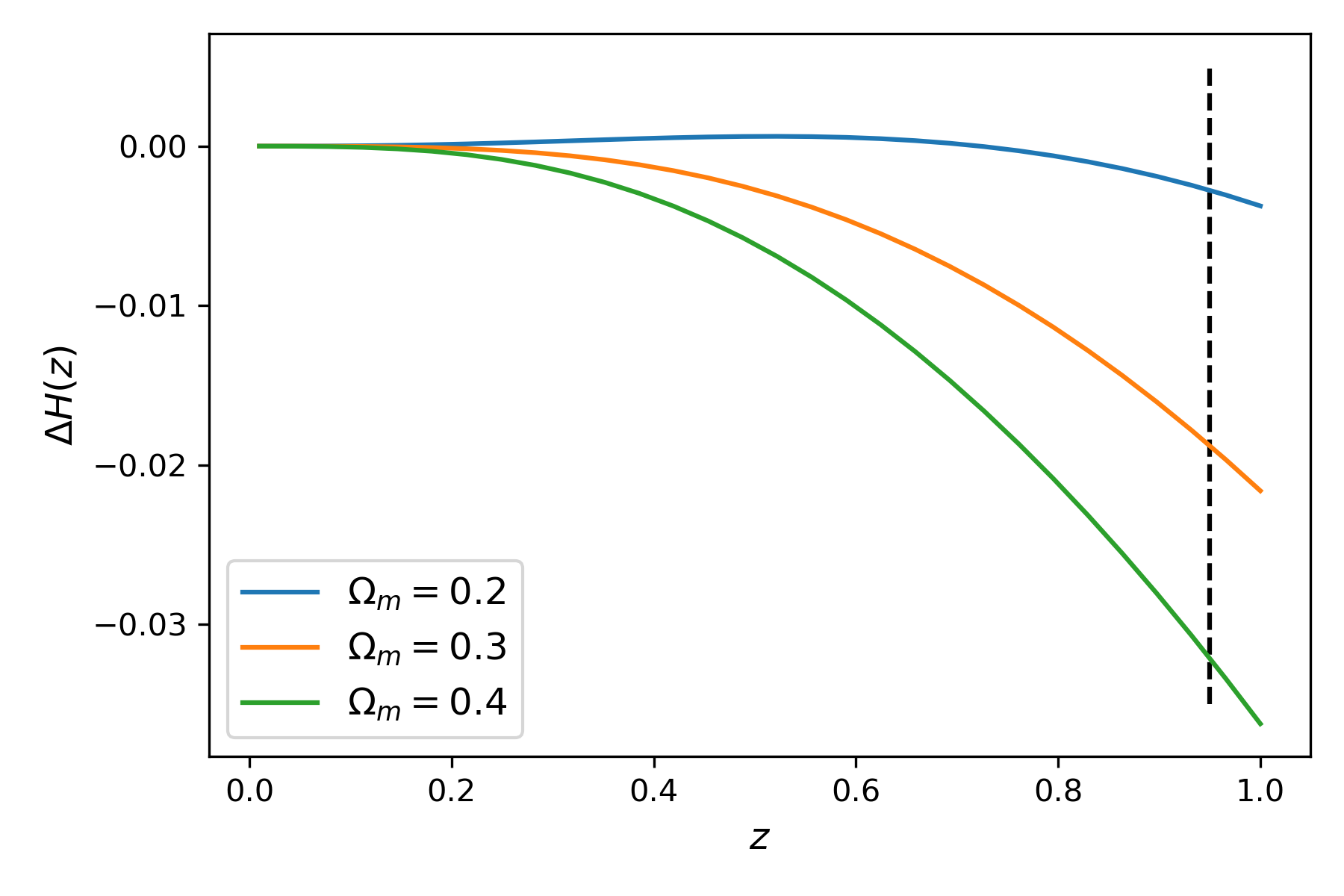}&
\includegraphics[width=75mm]{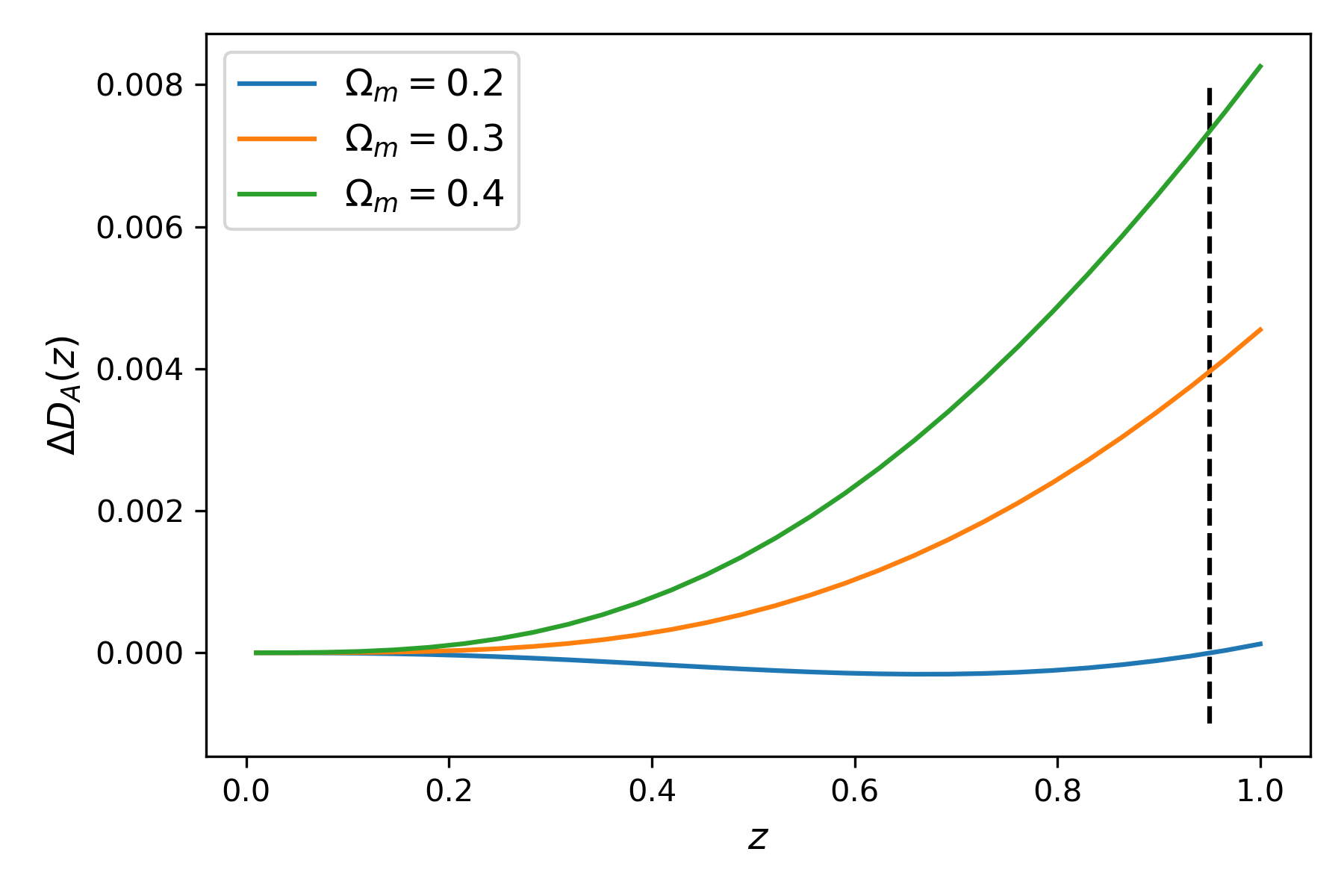}
\end{tabular}
\caption{Fractional error in $H(z)$ and $D_{A}(z)$ approximations at second order. Dashed lines denote the highest redshift in our sample.}
\label{3rdapprox} 
\end{figure}

\end{document}